\documentclass[referee,usegraphicx]{mn2e}
\title{Star formation in bright-rimmed clouds and cluster associated with W5 E H{\sc ii} region}
\author[Chauhan et al.]
{Neelam Chauhan$^{1}$\thanks{E-mail: neelam@aries.ernet.in}, A. K. Pandey$^{1}$, K. Ogura$^{2}$, J. Jose$^{1}$, D. K. Ojha$^3$,  
\newauthor M. R. Samal$^1$, H. Mito$^4$ \\
$^1$Aryabhatta Research Institute of observational sciencES (ARIES), Nainital, 263 129, India\\
$^2$Kokugakuin University, Higashi, Shibuya-ku, Tokyo 150-8440, Japan\\
$^3$Tata Institute of Fundamental Research, Mumbai (Bombay) - 400 005, India\\
$^4$Kiso Observatory, School of Science, University of Tokyo, Mitake,  Kiso-machi, Kiso-gun, Nagano-ken 397-0101, Japan\\
}

\begin{document}

\date{}

\pubyear{2010}

\maketitle

\label{firstpage}

\begin{abstract}
The aim of this paper is to present the results of photometric investigations 
of the central cluster of the W5 E region as well as a follow-up study of 
the triggered star formation in and around bright-rimmed clouds (BRCs). We 
have carried out wide field $UBVI_c$ and deep $VI_c$ photometry of the W5 E H{\sc ii} 
region.  A distance of $\sim$2.1 kpc and a mean age of $\sim$1.3 Myr have been obtained 
for the central cluster. The young stellar objects (YSOs) associated with the region 
are identified on the basis of near-infrared and mid-infrared 
observations. We confirmed our earlier results that the 
average age of the YSOs lying on/inside the rim are younger than those 
lying outside the rim. The global distribution of the YSOs shows an aligned 
distribution from the ionising source to the BRCs. These facts indicate 
that a series of radiation driven implosion processes proceeded 
from near the central ionising source towards the periphery of the W5 E 
H{\sc ii} region. We found that, in general, the age distributions of  the Class II  and 
Class III sources are the same. This result is apparently in 
contradiction with the conclusion by Bertout, Siess \& Cabrit (2007) and 
Chauhan et al. (2009) that classical T Tauri stars evolve to weak-line 
T Tauri stars. The initial mass function of the central cluster region 
in the mass range $0.4 \le M/M_\odot \le 30$ can be represented by 
$\Gamma = -1.29  \pm  0.03$. The cumulative mass functions indicate that 
in the mass range  $0.2 \le M/M_\odot \le 0.8$, the cluster region and BRC NW 
have more low mass YSOs in comparison to BRCs 13 and 14.

\end{abstract}


\section{Introduction}
 Massive stars have a profound effect on the evolution of 
their natal molecular clouds. 
Their strong stellar winds and ultraviolet (UV) radiation cause an important feedback of energy 
and momentum in the surrounding medium. Once massive stars form they begin to ionise 
the remaining parental molecular cloud. The ionising UV radiation has two competing effects on the 
parental molecular cloud. One is negative feedback on star formation activity, i.e., the remaining 
molecular cloud is dispersed and further star formation is halted. The other is 
positive feedback. The interaction can trigger new episodes of star formation as the H{\sc ii} 
region expands into the molecular cloud. Two processes have been considered 
for the triggering of star formation at the edge of the H{\sc ii} region, namely `collect and collapse' and 
`radiation driven implosion (RDI)'. 
In the collect and collapse process, a compressed layer of high density neutral material is 
accumulated between the 
ionisation front (IF) and the shock front (SF). The dynamical 
instabilities in the compressed layer result in the fragmentation of the layer and formation of 
second generation stars.

If the IF/ SF encounters pre-existing denser parts, it compresses them to induce star 
formation. The process is known as RDI.  Detailed model 
calculations of the RDI process have been carried out by several authors (e.g., Bertoldi 1989, 
Lefloch $\&$ Lazareff 1995, Kessel-Deynet \& Burkert 2003, Miao et al. 2006). Star formation is 
predicted to occur in {\bf the} initial short compression phase. This compression phase is followed by 
a transient phase of re-expansion and then by a quasi-stationary cometary phase. During this 
last phase, the cloud represents a structure of a dense head and a long tail and moves slowly 
away from the ionising source by the rocket effect. The signature of the RDI 
process is the anisotropic density distribution of gas in a relatively small molecular cloud 
surrounded by a curved ionisation front (bright rim) as well as a small group of YSOs in 
front of it.  
 
BRCs are small molecular clouds located near the edge of 
evolved H{\sc ii} regions and show signs of the RDI and are hence considered to be 
good laboratories to study the physical processes involved in the
RDI process. Sugitani et al. (1995) carried out near infrared (NIR) imaging of 44 BRCs and found 
that elongated small clusters or aggregates of YSOs, which are aligned toward the direction 
of the ionizing star, are often associated with them. These aggregates showed a tendency 
that `redder' (presumably younger) stars tend to be located inside the BRCs, whereas 
relatively `bluer' (presumably older) stars are found outside the clouds, suggesting an 
age gradient. Thus, they advocated a hypothesis called `{\it small-scale sequential star 
formation} ({\it S$^4$F})'. If the BRC is originally relatively large, the star formation may propagate along the axis of the BRCs 
as the ionization/shock front advances further and further into the molecular cloud (Kessel-Deynet \& Burkert 2003). 

The W5 H{\sc ii} region is an extended H{\sc ii} region with relatively simple morphology 
and  shows indications of triggered star formation. It is  a part of the large W3/W4/W5 
cloud complex in the Perseus arm and consists of two adjacent circular H{\sc ii} regions, 
W5 E and W5 W. 
There are many studies on this region. Karr \& Martin (2003) discussed triggered star 
formation in W5 using multi-wavelength archival data. Based on the timescales of the expansion 
of the H{\sc ii} region and the age of the YSOs, they obtained 
the timescale of the interaction between the molecular clouds and the H{\sc ii} region, $t \sim 0.5-1.0$ Myr.
 Using the $Spitzer$ Space Telescope imaging with Infrared Array Camera (IRAC) and Multiband 
Imaging Photometer for $Spitzer$ (MIPS), Koenig et al. (2008) noticed 
dense clusters of YSOs, centered around the O stars HD 17505, HD 17520, 
BD +60 586 and HD 18326. The H{\sc ii} region W5 E is primarily ionised by HD 18326. 
Chauhan et al. (2009) also paid attention to the cluster around this O7 V star. 
W5 E H{\sc ii} region has two 
BRCs, namely BRCs 13 and 14 (Sugitani et al. 1991, hereafter SFO 91) at its periphery. Based on 
the column densities of $^{13}$CO and the spatial distribution of YSO candidates, Niwa et al. (2009) 
identified a BRC candidate in the north-western part of the W5 E H{\sc ii} region. We refer 
to this BRC candidate as BRC NW. 
Hence W5 E is an interesting region to study triggered star formation.

In our earlier papers (Ogura et al. 2007, hereafter Paper I; Chauhan et al. 2009, hereafter Paper II) we have studied the star formation scenario in/ around 
six BRCs, including BRCs 13 and 14. The analysis was limited to the YSOs detected  using the 
H$\alpha$ emission and NIR excess only. A recent study by Koenig et al. (2008) has 
increased the number of YSOs in W5 E significantly. Hence, these data can be used to 
further investigate the YSO contents which we partly failed to detect in the H$\alpha$ and NIR 
excess surveys reported in Paper I and Paper II.

In this paper, we have made an attempt to make photometric studies of the stellar content of 
the newly identified cluster in W5 E, by incorporating the NIR and mid infrared (MIR) data from  IRAC/ MIPS of 
the $Spitzer$ telescope. We have also considered the properties of the YSOs in the 
cluster as well as the three BRC regions to understand the star formation scenario in 
the W5 E H{\sc ii} region. The influence of the ionizing source on the 
surrounding parental molecular cloud has also been discussed. 

In Sections 2 and 3, we describe the observations, data reductions and archival 
data used in the present work. 
Section 4 describes the analysis of the associated cluster. Sections 5 and 6 illustrate 
the procedure to estimate the membership, age, mass of the YSOs  and total-to-selective extinction, $R_V$ in the cluster 
and BRC regions. In Section 7 we discuss the initial mass function (IMF) and cumulative mass functions (CMFs) of the cluster and BRC regions. Section 8 describes the star formation scenario in the cluster and BRC regions. 
In Section 9 the disk evolution of T Tauri Stars (TTSs) has been discussed.
\section{Observations and Data reductions}
The 50$\times$50 arcmin$^2$ area containing the  cluster  around the O7 star 
HD 18326 that is noticed by Koenig et  al. (2008) and Chauhan et al. (2009), is reproduced from the DSS2-R image, 
and shown in Fig. 1. As is evident from the figure, the cluster  is embedded in an H{\sc ii} 
region W5 E. BRCs 13 and 14 are located towards the eastern side, whereas BRC NW detected  
by Niwa et al. (2009) can be seen towards the north-west direction of the cluster. 
In the ensuing subsections we describe the observations carried out to study the region in detail.

\subsection{Optical CCD Observations}

The $UBVI_c$ CCD optical observations of W5 E were carried out using the 
the 2048 $\times$ 2048 pixel$^2$ CCD camera mounted at the f/3.1 Schmidt focus of 
the 1.05-m telescope of Kiso Observatory, Japan. The pixel size of 
24 $\mu$m with an image scale of $1^{\prime\prime}.5$/pixel covers a field 
of view of $\sim$ 50$\times$ 50 arcmin$^2$ on the sky. The average FWHM of the star images during 
the observations was $\sim$$3^{\prime\prime}$.  
\begin{table}
\caption{Log of observations} 
\begin{tabular}{|p{.7in}|p{.5in}|p{3.in}|p{1.2in}|}
\hline
$\alpha_{(2000)}$ & $\delta_{(2000)}$ & Filter \& Exposure(sec)$\times$no. of frames& Date of observations\\
(h:m:s)           & (d:m:s)           &  &(yr-mm-dd)              \\
\hline
KISO&&&\\
02:59:22.60 &+60:33:48.7      & U:180$\times$9,30$\times$6; B:60$\times$9,10$\times$6; V:60$\times$9,10$\times$6; I:60$\times$9,10$\times$6 &2007-10-20\\
&&&\\
ST&&&\\
02:59:22.60 &+60:33:48.7      & U:240$\times$2,10$\times$1; B:180$\times$2,5$\times$1; V:100$\times$2,4$\times$1; I:60$\times$2,4$\times$1  & 2009-10-13\\
02:59:22.60 &+60:33:48.7      & U:900$\times$1,300$\times$1; B:600$\times$5; V:360$\times$9; I:120$\times$10 & 2006-12-16\\
03:00:08.34 &+60:39:19.3      & V:600$\times$5;I:300$\times$5 &  2007-10-13\\
02:58:38.78 &+60:39:06.1      & V:600$\times$5;I:300$\times$5 &  2007-10-13\\
03:00:03.49 &+60:27:59.6      & V:600$\times$5;I:300$\times$5 &  2007-10-13\\
02:58:33.34 &+60:28:47.2      & V:300$\times$10;I:300$\times$5 &  2007-11-06\\
02:57:44.41 &+60:38:29.1      & V:180$\times$2;I:100$\times$2 &  2009-10-13\\
02:57:44.41 &+60:38:29.1      & V:600$\times$6;I:300$\times$8 &  2009-10-15\\
02:59:34.13 &+61:05:28.4      & V:300$\times$10;I:300$\times$5 &  2006-12-15\\
&&&\\
HCT&&&\\
02:59:23.83 &+60:34:00.0      &Gr7/167l:300$\times$1 &2009-11-16\\

\hline
\hline
\end{tabular}
\label{tab1}
\end{table}

 The $UBVI_c$ CCD observations of the central region of  W5 E have also been carried out 
using the 2048 $\times$ 2048 
pixel$^2$ CCD camera mounted on the 1.04-m Sampurnanand Telescope (ST) of the 
Aryabhatta Research Institute of Observational Sciences (ARIES), Nainital, 
India. To improve the signal-to-noise ratio (S/N), the observations were 
carried out in a binning mode of $2 \times 2$ pixel$^2$. The pixel size is 24 $\mu$m  
with an image scale of  $0^{\prime\prime}.37$/pixel  and the entire chip 
covers a field of $\sim$13$\times$13 arcmin$^2$ on the sky. The average FWHM of star images 
was $\sim$$2^{\prime\prime}.5$.
The observations of the central region of W5 E were standardised on 2009 October 
13 by observing standard stars in the SA 92 field (Landolt 1992).
Deep imaging of a nearby field region towards the north from the 
cluster centre ($\alpha_{2000}$ = $02^{h}59^{m}34^{s}$; 
$\delta_{2000}$ = $+61^{\circ}05^{\prime}28^{\prime\prime}$), was also carried out in $V$ and $I_c$ 
bands using the ST. The magnitudes of bright stars which were saturated in deep exposure frames 
have been taken from short exposure frames.  
A number of bias and twilight frames were also taken during the observing runs. 
The log of the observations is tabulated  in Table \ref{tab1}. 

The pre-processing of the data frames was done using the various tasks 
available under the $IRAF$ data reduction software package. The photometric 
measurements of the stars were performed using the $DAOPHOT II$ 
software package (Stetson 1987). The point spread function was obtained for each 
frame using several uncontaminated stars.

The instrumental magnitudes of the central region observed on 2009 October 13 with ST 
were converted into the standard system using 
least-square linear regression procedures outlined by Stetson (1992). 
The photometric calibration equations used are as follows:\\
\noindent
 $u = U + (7.858 \pm 0.007) + (0.596 \pm 0.024) X + (0.124 \pm 0.008)(U-B)$,\\
\noindent
 $b = B + (5.464 \pm 0.006) + (0.336 \pm 0.010) X + (0.134 \pm 0.006)(B-V)$,\\
\noindent
 $v = V + (5.088 \pm 0.006) + (0.188 \pm 0.011) X + (0.032 \pm 0.007)(V-I_c)$,\\
\noindent
 $i = I_c + (5.320 \pm 0.012) + (0.121 \pm 0.019) X + (0.106 \pm 0.011)(V-I_c)$\\
where  $U, B, V$ and $I_c$ are the standard magnitudes; $u, b, v$ and $i$ are 
the instrumental magnitudes obtained after time and aperture corrections and 
X is the airmass. We have ignored the second-order colour correction terms 
as they are generally small in comparison with other errors present in the 
photometric data reduction. 
The standard deviations of the standardisation residuals, $\Delta$, between 
the standard and transformed magnitudes and colours of the standard stars, 
are found to be $\Delta$$V = 0.008$, $\Delta$$(B-V) = 0.017$, $\Delta$$(V-I_c) 
= 0.020$ and  $\Delta$$(U-B) = 0.011$.
The photometric accuracies depend on the brightness of the stars, and the typical $DAOPHOT$ errors 
in $B, V$ and $I_c$ bands at $V$ $\sim$ 18 are smaller than 0.01 mag. Near the limiting magnitude of 
$V$ $\sim$ 21,  the $DAOPHOT$ errors increase to 0.06 and 0.04 mag in the $V$ and $I_c$ bands, 
respectively. The Kiso data were standardised using the secondary standards obtained from the central region observations as mentioned above.

 To study the luminosity function (LF)/ mass function (MF) of the cluster region we have used $VI_c$ data taken with ST. 
It is necessary to take into account the 
incompleteness in the observed data that may occur for various reasons (e.g., crowding of the stars). 
A quantitative evaluation of the completeness of the photometric data with respect to the brightness 
and the position on a given frame is necessary to convert the observed LF 
to a true LF. We used the ADDSTAR routine of $DAOPHOT II$ to determine the completeness factor (CF). 
The procedure has been outlined in detail in our earlier works (see e.g., Pandey et al. 2001). 
We randomly added artificial stars to both $V$ and $I_c$ images in such a way that they 
have similar geometrical locations but differ in $I_c$ brightness according to the mean $(V-I_c)$ 
colour (=1.5 mag) of the data sample. The luminosity distribution of artificial stars is chosen 
in such a way that more stars are inserted towards the fainter magnitude bins. The frames are 
reduced using the same procedure used for the original frame. The ratio of the number of 
stars recovered to that added in each magnitude interval gives the CF as a function of magnitude. 
The minimum value of the CF of the pair (i.e., $V$ and $I_c$ band observations) for the two sub-regions, 
given in Table \ref{cf_opt}, is used to correct the data for incompleteness. The incompleteness of the data 
increases with increasing magnitude as expected. However, it does not depend on the area 
significantly.
\begin{table}
\caption{Completeness factor of photometric data in
the cluster and field regions.}
\label{cf_opt}
\begin{tabular}{cccc} \hline
V range&    cluster region&  & field region\\
 (mag)& $r\le 3^\prime$ & $3^\prime <r \le6^\prime$&  \\
\hline
13.5 - 14.0&1.00&1.00&1.00\\
14.0 - 14.5&1.00&1.00&1.00\\
14.5 - 15.0&0.99&0.99&1.00\\
15.0 - 15.5&0.98&0.98&0.99\\
15.5 - 16.0&0.98&0.98&0.98\\
16.0 - 16.5&0.96&0.97&0.98\\
16.5 - 17.0&0.97&0.96&0.98\\
17.0 - 17.5&0.97&0.96&0.98\\
17.5 - 18.0&0.95&0.95&0.96\\
18.0 - 18.5&0.93&0.94&0.95\\
18.5 - 19.0&0.91&0.92&0.94\\
19.0 - 19.5&0.88&0.90&0.91\\
19.5 - 20.0&0.85&0.84&0.89\\
20.0 - 20.5&0.81&0.82&0.84\\
20.5 - 21.0&0.72&0.71&0.77\\
21.0 - 21.5&0.51&0.54&0.57\\
\hline
\end{tabular}
\end{table}

The photometric results for the BRC 13 and BRC 14 regions have been taken from 
our earlier work (Paper II). They are based on observations with Himalayan 
Faint Object Spectrograph Camera (HFOSC) on the 2.0-m Himalayan Chandra Telescope 
(HCT). The boundaries of the earlier observations are shown with dashed lines in Fig. 1.

\subsection { Grism Slit spectroscopy} 
We obtained a low resolution optical spectrum of the exciting star of W5 E, HD 18326, on 2009 November 
16 using HFOSC on HCT, with a 
slit width of 2 arcsec and Grism 7 ($\lambda = 3800 - 6840$ 
\AA, dispersion = 1.45 \AA /pixel). A one dimensional spectrum was extracted from the bias subtracted 
and flat-field corrected image in the standard manner using IRAF. The wavelength calibration of the 
spectrum was done using a FeAr lamp source. The standard star G191-B2B is used for the 
standardisation and flux calibration.

\section{Archival Data}
\subsection { Near-infrared data from 2MASS}
NIR $JHK_s$ data for point sources within a radius of 
$25^\prime$ around the central cluster have been obtained from the Two Micron 
All Sky Survey (2MASS) Point Source Catalog (PSC) (Cutri et al. 2003). 
Sources having uncertainty less than 0.1 mag (S/N $\ge$ 10) in all the three 
bands were selected to ensure high quality data. The $JHK_s$ data were transformed
from the 2MASS system to the CIT system using 
the relations given at the 2MASS website\footnote{http://www.astro.caltech.edu/$\sim$jmc/2mass/v3/transformations/}.

\subsection { Mid-infrared data from $Spitzer$}
The near- and mid-infrared data (3.6 to 24 $\mu$m) from the $Spitzer$ Space telescope  have 
provided the capability to detect and measure 
the infrared excesses due to circumstellar disk emission of the YSOs. In order to study the 
evolutionary stages of the YSOs detected using the $Spitzer$, we used IRAC (3.6 $\mu$m, 
4.5 $\mu$m, 5.8 $\mu$m and 8.0 $\mu$m) and MIPS (24 $\mu$m) photometry taken from
 Koenig et al. (2008).
\subsection {H$\alpha$ emission stars from slitless spectroscopy}
 The H$\alpha$ emission stars for the cluster region and BRC regions have been taken from 
Nakano et al. (2008) and Ogura et al. (2002), respectively.

\section{Analysis of the associated cluster}
\subsection{Radial stellar surface density profile}\label{rden}

The radial extent is one of the important parameters to study the dynamical  
properties of clusters. To estimate this we assumed a spherically symmetric distribution of stars in 
the cluster. 
The star count technique is one of the useful tools to determine the distribution 
of cluster stars with respect to the surrounding stellar background.

In order to determine the cluster centre, we derived the highest peak of stellar density 
by fitting a Gaussian profile to the star counts in strips along both 
the X and Y axes around the eye estimated cluster centre. The cluster centre from the 
optical data has turned out to be at $\alpha_{2000}$ = $02^{h}59^{m}22^{s}.0$$\pm$$ 1^{s}.0$; 
$\delta_{2000}$ =
$+60^{\circ}34^{\prime}37^{\prime\prime}$$\pm 12^{\prime\prime}$. 
 We repeated the same procedure using the 2MASS data to estimate the cluster 
centre and obtained it to be $\sim$$12^{\prime\prime}$ away from
the optical co-ordinates. However, this difference is within the uncertainty. 
Henceforth, we adopt the optical centre.

We estimated the radial density profile (RDP) to study the radial structure of the cluster. 
We divided the cluster into a number of concentric 
circles and the projected stellar density in each concentric annulus was obtained by dividing the 
number of stars by the respective annulus area. Stars brighter than $V$ = 19.5 mag and $K$ =  14.7 mag were 
considered for estimating the RDPs from the optical and 2MASS data, respectively.
The densities thus obtained are plotted as a function of radius in Fig. 2. 
The  error bars are derived assuming that the number of stars in each annulus follows the Poisson 
statistics.

The extent of the cluster $r_{cl}$ is defined as the radius where the cluster stellar 
density merges with the field stellar density. The horizontal dashed line in the plot 
shows the field star density, which is obtained from a region $\sim$ 25$^\prime$ away 
towards the north from the cluster centre ($\alpha_{2000}$ = $02^{h}59^{m}34^{s}$; 
$\delta_{2000}$ = $+61^{\circ}05^{\prime}28^{\prime\prime}$). Based on the  radial 
density profile, we find that the $r_{cl}$
 is about 6$^\prime$ for stars brighter than $V$ = 19.5 mag. Almost the same value for 
the cluster extent is obtained for the 2MASS data. We adopted a radius of 6$^\prime$ for this 
cluster to obtain the cluster parameters such as reddening, 
distance, IMF etc.

To parameterise the RDP of the cluster, we fitted the observed RDP with the 
empirical model of King (1962) which is given by
\begin{center}

$\rho (r) \propto {\displaystyle{\rho_0} \over \displaystyle{1+\left({r\over r_c}\right)^2}}$,  

\end{center}
where $r_c$ is the core radius at which the surface density $\rho(r)$ becomes half of the central 
density, $\rho_0$ . The best fit to the radial density obtained by a $\chi^2$  minimization 
technique, is shown in Fig. 2. The core radius thus estimated is 1$^\prime$.01 $\pm$ 0$^\prime$.12. 
 
\subsection{Interstellar reddening}
We know that the extinction or reddening $E(B - V)$ of a star in a cluster arises due to two 
distinct sources: 
\begin{enumerate}
\item the general interstellar medium (ISM) in the foreground of the cluster [$E(B −- V)_{f}$], and 
\item the localized ISM associated with the cluster [$E(B −- V )_c$ = $E(B −- V)$ $−-$ $E(B −- V)_{f}$], 
\end{enumerate}
The former component is 
characterised by the ratio of the total-to-selective extinction $R_V$ [= $A_V$ /$E(B −- V )$] = 3.1  
(Wegner 1993; He et al. 1995; Winkler 1997), whereas, for the intra-cluster regions of young clusters 
embedded in a dust and gas cloud, the value of $R_V$ may vary significantly (Chini \& Wargau 1990; Tapia 
et al. 1991; Pandey et al. 2000). The value of $R_V$ affects the distance determination significantly, and consequently the 
age determination of stars. Several studies have already pointed out {\bf an} anomalous 
reddening law with high $R_V$ values in the vicinity of star forming regions (see e.g., Neckel \& Chini 
1981, Chini \& Kr{\"u}egel 1983, Chini \& Wargau 1990, Pandey et al. 2000, 
Samal et al. 2007). Since the W5 E cluster  and the BRCs are associated with the H{\sc ii} region, 
it will be interesting to examine the reddening law in these objects. The ratio of total-to-selective 
extinction $R_V$ is found to be normal in the cluster region (cf. Section 6 for details).

Since spectroscopic observations are not available, the interstellar reddening $E(B −- V)$ towards the 
cluster region is estimated using the $(U −- B)$/$(B −- V)$ colour-colour (CC) diagram. The CC diagram 
of the cluster region  is presented in Fig. 3. Since the cluster is very young, a variable reddening 
within the cluster region is expected. In Fig. 3, the continuous lines represent the intrinsic 
zero age main sequence (ZAMS) by Girardi et al. (2002) which are shifted by $E(B −- V )$ = 0.62 
and 0.80 mag respectively, along the normal reddening vector (i.e., $E(U −- B)$/$E(B −- V)$ = 0.72)  
to match the distributions of probable cluster members. Fig. 3 thus yields a variable reddening 
with $E(B −- V)_{min}$ = 0.62 mag to $E(B −- V)_{max}$ = 0.80 mag in the cluster region. 
The stars lying within these two reddened ZAMSs may be probable members of the cluster. 
Reddening of individual stars with spectral classes earlier than A0 have been computed using 
the reddening free index, Q (Johnson \& Morgan 1953). Assuming a normal reddening 
law, we calculated $Q = (U - B) - 0.72 \times (B - V)$. 
The value of Q for stars earlier than A0 will be $<$ 0. For main sequence (MS) stars, the intrinsic $(B - V)_0$ 
colour and colour excess can be obtained by the relations; $(B - V)_0 = 0.332 \times Q$ 
and $E(B - V) = (B - V) - (B - V)_0$, respectively.  
Fig. 3 also indicates a large amount of contamination due to field stars. The probable 
late type foreground stars with spectral types later than A0 may follow the ZAMS 
reddened by $E(B −- V )=0.50$ mag. A careful 
inspection of Fig. 3 indicates the presence of further reddened background 
populations.  The reddening  $E(B −- V)$ for the background population is found out to be  
in the range of $\sim$ 0.95  -  1.25 mag. This population may belong to the blue plume (BP) 
of the Norma-Cygnus arm (cf. Pandey et al. 2006). The estimated  $E(B −- V)$ 
values for the background population are comparable to the $E(B −- V)$ value of the BP population 
around $l \sim 130 ^{\circ}$ (cf. Pandey et al. 2006).

\subsection{Spectral classification of the ionising star in W5 E H{\sc ii} region}
 We obtained a slit spectrum of the brightest source to study its nature. In Fig. 4, we present the flux calibrated, normalised spectrum of HD 18326, which is the ionising source of the H{\sc ii} region, in the wavelength range 
$3990 ­- 5000$ \AA. The important lines have been identified and labeled. HD 18326 is identified 
as an O7V and O7V(n) type star by Conti \& Leep (1974) and Walborn (1973), respectively. The ratio of HeI $\lambda$4471/
HeII $\lambda$4542 is a primary indicator of the spectral class of early type stars. The 
ratio we found for this star is $\sim$1, which indicates that it is an O7 $\pm$ 0.5 star. 
With the present resolution of the spectrum, luminosity assessment is quite difficult. However, due to  
the presence of strong absorption in HeII $\lambda$4686, we assign the luminosity class V.

 \subsection{Distance and optical colour magnitude diagrams}

The spectral class of the ionising source yields an intrinsic distance modulus of 11.2 
which corresponds to a distance of 1.74 kpc. Here, it is worthwhile noting that 
the $M_V$ for an O7V star in the literature varies significantly; e.g., $M_V$ = -5.2 
(Schmidt-Kaler 1982) to -4.9 (Martins \& Plez 2006). This star is also reported as a 
variable star and a suspected spectroscopic binary (Kazarovets et al. 1998; 
Turner et al. 2008). Hence, the distance estimation 
based only on the O-type star alone may not be reliable. We also estimated the individual 
distance modulus of other probable MS stars. The intrinsic colours for each star were estimated 
using the Q-method as discussed in Sec. 4.2. We estimated corresponding $M_V$ values for each 
star using the ZAMS by Girardi et al. (2002). The average value of the intrinsic distance 
modulus, obtained using 24 probable MS members, comes out to be $11.65 \pm 0.57$, 
corresponding to a distance of $2.1 \pm 0.3$ kpc.    
 This distance estimate is in agreement with those obtained by Becker \& 
Fenkart (1971; 2.2 kpc), Georgelin \& Georgelin (1976; 2.0 kpc) and 
Hillwig et al. (2006; 1.9 kpc).

We used the optical colour magnitude diagrams (CMDs) to derive the fundamental 
parameters of the cluster, such as age, distance etc. The $V/(B −- V )$ and $V/ (V −- I )$ CMDs for stars lying 
within 6$^\prime$ radius are shown in Figs. 5a and 5b. Fig. 5c shows the $V/ (V −- I )$ CMD for a 
nearby field region (see Sec. 4.1). 
A 4 Myr isochrone for $Z$ = 0.019 by Girardi et al. (2002) and the 
pre-main-sequence (PMS) isochrones for 1 Myr and 5 Myr by Siess et al. 
(2000) have also 
been plotted for  $(m - M_V)$ =  13.5 mag and $E(B - V)$ = 0.62 mag assuming $E(V - I)$ = 1.25$\times$$E(B - V)$ and $R_V = 3.1$. 
A comparison of the CMDs of the cluster region with the field region reveals an unambiguous population of 
PMS sources along with a significant contamination due to field star population.

\section{IDENTIFICATION OF PRE-MAIN-SEQUENCE OBJECTS ASSOCIATED WITH THE CLUSTER AND BRCS}
Since the W5 E region is located at a low galactic latitude, the region is 
significantly 
contaminated by foreground/background stars as discussed above. In order to understand 
star formation in the region, we selected probable PMS members associated with the region 
using the following criteria.

Some PMS stars, specifically classical T Tauri stars (CTTSs), show emission lines in their spectra, among 
which usually H$\alpha$ is the strongest. Therefore, H$\alpha$ emission stars can be considered as 
good candidates for PMS stars associated with the region. In the present study, we use H$\alpha$ emission 
stars found by Ogura et al. (2002) and Nakano et al. (2008) in the W5 E H{\sc ii} region. Since many PMS 
stars also show NIR/MIR excesses caused by circumstellar disks, NIR/MIR photometric surveys are also 
powerful tools to detect low-mass PMS stars. We have also used the YSOs identified 
by Koenig et al. (2008) in the W5 E H{\sc ii} region using $Spitzer$ IRAC and MIPS photometry. 

 Fig. 6a shows the NIR ${(J - H)/(H - K)}$ colour-colour (NIR-CC) diagram of all the 
sources detected in the 2MASS catalogue along with the YSOs identified by Koenig et al. (2008) in the cluster region, whereas 
Fig. 6b shows the NIR-CC diagram for the sources in the nearby reference field. In Figs. 6a and 6b, the thin and 
long-dashed curves represent the unreddened MS and giant branches (Bessell $\&$ 
Brett 1988), respectively. The dotted line indicates the locus of intrinsic CTTSs (Meyer et al. 
1997). The curves are also in the CIT system.  The parallel dashed lines are the reddening 
vectors drawn from the tip (spectral type M4) of the giant branch (``upper reddening line''), 
from the base (spectral type A0) of the MS branch (``middle reddening line'') 
and from the tip of the intrinsic CTTS line (``lower reddening line''). 
The extinction ratios $A_J/A_V = 0.265, A_H/A_V = 0.155$ and $A_K/A_V=0.090$ have been 
adopted from Cohen et al. (1981). We classified sources into three regions in the NIR-CC 
diagrams (cf. Ojha et al. 2004a). `F' sources are located between the upper and middle reddening 
lines and are considered to be either field stars (MS stars, giants) or 
Class III sources and Class II sources with small NIR excesses. `T' sources are located between 
the middle and lower reddening lines.  These sources are considered to be mostly CTTSs 
(Class II objects). There may be an overlap in NIR colours of Herbig Ae/Be stars and 
CTTSs in the `T' region (Hillenbrand et al. 1992). `P' sources are those located in 
the region redward of the `T' region and are most likely Class I objects 
(protostar-like objects; Ojha et al. 2004b). So, objects falling in the `T' and `P'  regions of 
NIR-CC diagrams are considered to be NIR excess stars and hence are probable members of the 
cluster. These sources are included in the analysis of the  present study in addition 
to H$\alpha$ emission stars. It is worthwhile, however, to note that Robitaille et al. 
(2006) have recently shown 
that there is a significant overlap between protostars and CTTSs in the NIR-CC space.

 The $V/(V - I)$ CMD for the YSOs taken from the catalogue by Koenig et al. (2008) for the cluster 
region ($r_{cl} = 6^\prime$) is shown in Fig. 7. Here the well-known age-with-mass trend that 
higher mass stars look older than lower mass stars (Hillenbrand et al. 2008) is evident. In 
addition a few sources, having $V \ga 15$, classified as Class III objects are located near 
the MS.
 They are also located on the 
MS in the NIR-CC diagram and hence could be field stars. This indicates that 
part of Class III objects by Koenig et al. (2008) are not YSOs, but rather stars 
found in the W5 Spitzer photometric sample that appear to have photospheric colours in 
the 3 to 24 $\mu$m bands  and thus appear as Class III SEDs.  Thus, the listed Class III 
objects by Koenig et al. (2008) may be heavily contaminated by foreground and background field populations. 
A  comparison of the NIR-CC diagram of the cluster region and nearby field region indicates that 
the YSOs identified by Koenig et al. (2008) and having $(J-H) > 0.7$ mag and $(H-K) > 0.3$ 
mag can be safely considered as YSOs associated with the region.

The NIR-CC diagrams for the identified YSOs (with H$\alpha$ emission and IR excess) in the 
BRC NW, BRC 13 and BRC 14 are shown in Figs. 8a, 8b and 8c, respectively. The NIR-CC diagrams were used 
to estimate $A_V$ for each of these YSOs by tracing them back to the intrinsic CTTS
locus of Meyer et al. (1997) along the reddening vector (for details see Paper I \& Paper II).
The  $A_V$ for the stars lying in the `F' region is estimated by tracing them
back to the extension of the intrinsic CTTS locus. The mean reddening for each {\bf BRC} region is calculated. 
The mean $A_V$ values for 
BRC NW, BRC 13 and BRC 14  are found to be 2.26 mag, 2.33 mag and 3.05 mag, respectively. 
Fig. 9 shows the $V/(V-I)$ CMDs of the YSOs selected using the 
NIR-CC diagram in the cluster region as well as in the three BRC regions. Again, the CMDs indicate that 
practically all of them are PMS stars but, at the same time reveal a 
significant scatter in their age. The age of each YSO was estimated by referring to the 
isochrones. The mass of the YSOs was  estimated using the $V/(V −- I_c )$ colour-magnitude 
diagram as discussed in Pandey et al. (2008) and Chauhan et al. (2009).  The resultant  
age and mass are given in Table 3, which is available in an electronic form.  Here, we would 
like to point out that the estimation of the age of the PMS stars by comparing the observations 
with the theoretical isochrones is prone to two kinds of errors; random errors in observations and 
systematic errors due to the variation between the predictions of different theoretical evolutionary 
tracks (see e.g., Hillenbrand 2005, Hillenbrand et al. 2008). The effect of random errors in  
determination of  age and mass was estimated by propagating the random errors to their observed 
estimation by assuming normal error distribution and using Monte-Carlo simulations. 
The use of different PMS evolutionary models give different ages and hence an age spread in a cluster (e.g., Sung et al. 2000). 
In the present study, we have used the models by Siess et al. (2000) only for all the BRCs and 
the cluster region, therefore our age and mass estimations are not affected by the systematic errors. 
However, the use of different sets of PMS evolutionary tracks will introduce a systematic shift 
in age determination. The presence of binaries may be another source of errors in 
the age determination. Binarity will brighten the star, consequently the CMD will 
yield a lower age estimate. In the case of equal mass binaries we expect an error of $\sim$ 50 - 60\% 
in the age estimation of the PMS stars. However, it is difficult to estimate the influence of 
binarity on the mean age estimation as the fraction of binaries is not known.

\begin{table*}
\caption{The {\it B, V} and {\it I$_c$} photometric data along with their position, mass and age for the YSOs in the cluster and BRC regions}
\label{phot_data}
\begin{tabular}{cccccccc}
\hline
S. No.&$\alpha_{2000}$&$\delta_{2000}$&$B \pm eB$&$V \pm eV$&$I \pm eI$&$Age\pm \sigma$&$mass \pm \sigma$   \\
  &$h:m:s$&$d:m:s$&$mag$&$mag$&$mag$&$Myr$&$M_\odot$   \\
\hline
{\bf cluster region}&&&&&&&\\
1 &$02:58:39.03$&$60:37:25.8$&$-$&$19.67 \pm 0.02$&$16.78 \pm 0.02$&$0.7 \pm 0.01$&$0.4 \pm 0.00$   \\
2 &$02:58:39.24$&$60:37:02.7$&$20.07 \pm 0.02$&$18.21 \pm 0.01$&$15.60 \pm 0.01$&$0.2 \pm 0.02$&$0.5 \pm 0.00$   \\
3 &$02:58:39.32$&$60:35:00.6$&$-$&$20.78 \pm 0.03$&$18.05 \pm 0.02$&$2.4 \pm 0.05$&$0.4 \pm 0.01$   \\
-- &$------$&$------$&$-----$&$------$&$------$&$----$&$----$\\
-- &$------$&$------$&$-----$&$------$&$------$&$----$&$----$\\
\hline
\end{tabular}
\end{table*}
  
\section {Reddening law in the cluster and BRC regions}
\label{reddening}
To study the nature of the extinction law in the region, we used two colour diagrams (TCDs) as 
described by Pandey et al. (2003). The TCDs of the form of $(V −- \lambda)$ vs. $(B −- V )$, 
where $\lambda$ is one of the broad-band filters $(R, I, J, H, K, L)$, provide an effective  
method of separating the influence of the normal extinction produced by the diffuse 
interstellar medium from that of the abnormal extinction arising within regions having 
a peculiar distribution of dust sizes (cf. Chini \& Wargau 1990, Pandey et al. 2000).

The TCDs for the Class III YSOs associated with the cluster and BRC regions are shown in Figs. 10a - 10d. 
In order to avoid IR excess stars, we have used all the $Spitzer$ Class III sources having $V$ $<$ 
17 mag. The slopes of the distributions are given in the figure. The ratio of total-to-selective 
extinction `$R_V$' in the regions is estimated using the 
procedure described in Pandey et al. (2003). 
The $R_V$ values in the cluster and BRC regions, i.e., BRC NW, BRC 13 and BRC 14 are estimated to 
be $R_{cluster}$ = 3.14 $\pm$ 0.12, $R_{BRC  NW}$ = 3.46 $\pm$ 0.20, $R_{BRC 13}$ = 3.41 $\pm$ 0.03 and $R_{BRC 14}$ = 2.75 $\pm$ 0.12, respectively. 
The value of $R_{cluster}$
indicates a normal reddening law in the cluster region. The higher values of $R_{BRC NW}$ and $R_{BRC 13}$ indicate a larger 
grain size in the BRC NW and BRC 13 regions, whereas the smaller value of $R_{BRC 14}$ suggests a smaller grain size in the case of BRC 14. This indicates that the evolution of dust grains in the W5 E region 
has not taken place in a homogeneous way.

\section{Mass functions of the cluster and BRC regions}
\subsection{Initial mass function of the cluster}
The IMF is the distribution of the mass of stars 
at the time of a star formation event. Young clusters are preferred sites for IMF studies as 
their mass functions (MFs) can be considered as IMFs, since they are too young to lose 
significant number of members either due to dynamical evolution or stellar evolution. 
The IMF is defined as the number of stars per unit logarithmic mass interval, 
and is generally represented by a power law having a slope, \\ 
$\Gamma$ = {{ $\rm {d}$  log $N$(log $m$)}/{$\rm{d}$ log $m$}},\\
 where $N$(log $m$) 
is the  number of stars per  unit logarithmic mass interval.  For the mass range 
$0.4  < M/M_{\odot} \le 10$,  the classical  value derived  by Salpeter  (1955) 
is $\Gamma = -1.35$ .

To study the MF and LF, it is necessary to eliminate the field star contamination from the 
cluster region. In the absence of proper motion data, one can use statistical criteria to 
estimate the number of probable member stars in the cluster region. 
To remove the contamination due to field stars from the MS and PMS sample, we statistically
subtracted the contribution of field stars from the observed CMD of the cluster region using
the following procedure. For any star in the $V /(V - I)$ CMD of the control field (Fig. 5c), the
nearest star in the cluster's  $V/(V - I)$ CMD (Fig. 5b) within $V \pm 0.125$ and 
$(V - I) \pm 0.065$ was removed. The statistically cleaned CMD is shown in Fig. 11, 
which clearly shows a sequence of PMS stars. The PMS isochrones by Siess et al. (2000) 
for age of 0.1 and 5 Myr (dashed lines) and the 4 Myr isochrone by Girardi et al. (2002) 
(continuous line) are shown in Fig. 11. The evolutionary tracks by Siess et al. (2000) 
for different masses are also shown which are used to estimate the mass of 
PMS stars. 
Here we would like to {\bf note} that the points shown by filled circles in Fig. 11 may not represent the actual members 
of the clusters. However, the filled circles should represent the statistics of PMS stars 
that can be used to study the MF of the cluster region. We used the statistics 
of the sources having age 0.1 $\le$ age $\le$ 5 Myr in the statistically cleaned CMD (Fig. 11) 
to study the IMF of the cluster region. It is also important to make corrections 
in the data sample for incompleteness which may be due to various reasons, e.g., 
crowding of the stars.  We determined 
the CF as described in Sec. 2.1. 
Since the MS age of the most massive star in the cluster is $<$ 4 Myr, 
the stars having $V$ $<$ 14.7 mag (M $> 3 M$$_\odot$) have been considered on the MS. In order to obtain the MF 
for MS stars, the LF is converted to MF using the theoretical model of 
Girardi et al. (2002). The MF of the PMS stars was obtained by counting the 
number of stars in different mass bins (for details see Pandey et al. 2008, 
Jose et al. 2008). Fig. 12 shows the MF of the cluster in the mass range 
0.4 $\le$ M/$M_\odot$ $\le$ 30. A single slope for the MF in the mass range 
0.4 $\le$ M/$M_\odot$ $\le$ 30 can be fitted with $\Gamma$ = -1.29 $\pm$ 0.03, which 
is comparable to the  Salpeter value (-1.35).

\subsection{Cummulative mass function of identified YSOs in the cluster and BRC regions}

The MF is an important tool to compare the star formation process/scenario in 
different regions. Morgan et al. (2008), based on SCUBA observations, have estimated 
the mass of 47 dense cores within the heads of 44 BRCs. They concluded that 
the slope of the MF of these cores is significantly shallower than 
that of the Salpeter MF. They also concluded that it depends on 
the morphological type of BRCs (for the morphological description of BRCs 
we refer to SFO 91): `A' type BRCs appear to follow the mass spectrum of the 
clumps in the Orion B molecular cloud, whereas the BRCs of the `B' and `C' 
types have a significantly shallower MF. 

In Paper II, we have studied the CMFs of YSOs selected 
on the basis of NIR excess and H$\alpha$ emission and found that 
the `A' type BRCs seem to follow a MF similar to that found 
in young open clusters, whereas `B/C' type BRCs have a significantly 
steeper CMF, indicating that BRCs of the latter type tend to form relatively
low mass YSOs in the mass range 0.2 $\le$ M/M$_\odot$ $\le$ 0.8. The CMF 
of the YSOs associated with the `A' type BRCs in the mass range 
0.2 $\le$ M/M$_\odot$ $\le$ 0.8 is found to be comparable with the CMF 
of the average Galactic IMF (cf. Paper II).
\begin{table}
\caption{Cumulative mass function of the YSOs in the cluster and BRC regions}
\label{mfs}
\begin{tabular}{ccc} \hline
Region &CMF in the mass range &   \\
 & $0.8\le M/M_\odot \le 1.2$ &$0.2\le M/M_\odot \le 0.8$  \\
\hline
Cluster &-2.54 $\pm$ 0.40 & -1.97 $\pm$ 0.28\\
BRC NW  &-5.06 $\pm$ 3.18 & -2.63 $\pm$ 0.21\\
BRC 13  &-6.45 $\pm$ 2.01 & -0.98 $\pm$ 0.14\\
BRC 14  &-4.69 $\pm$ 0.75 & -1.08 $\pm$ 0.14\\ 
\hline
\end{tabular}
\end{table}

As mentioned above, in Paper II, the YSOs were selected on the basis of NIR 
excess and H$\alpha$ emission. To make the sample statistically 
significant we combined the data of different types of BRCs lying 
in different star forming regions. Since star formation and evolution of 
TTSs in and around BRCs may depend on the prevailing conditions in the region, 
it will be worthwhile to re-investigate the CMF of the YSOs (identified in 
the present work) in BRC NW, BRCs 13 and 14 as well as those 
associated with the cluster region.
Fig. 13 compares the CMFs of the YSOs in the BRC regions and the cluster region 
presuming that the biases (if any) in all the four samples are the same. 
It is interesting to note that the CMFs of the 
three BRCs show a break in the slope at $\sim$ 0.8 $M_\odot$. 
The slopes of the CMFs are given in Table 4, 
which indicate that in the case of three BRCs, the slopes  
in the mass 
range 0.8 $\le$ M/$M_\odot$ $\le$ 1.2 are almost the same. However, 
in the mass range $0.2\le M/M_\odot \le 0.8$, the slopes in the BRCs 13 
and 14 are found to be similar, whereas 
CMF in BRC NW and the central cluster are biased towards lower mass in 
comparison to the BRCs 13 and 14. The average Galactic 
IMF in the mass range $0.2\le M/M_\odot \le 0.6$, i.e., $\Gamma = -0.3$ 
(Kroupa 2001, 2002) corresponds to the slope of the CMF of -1.1 $\pm$ 0.1. 
This fact also indicates that the cluster region and BRC NW 
have relatively more low mass YSOs in the mass range $0.2\le M/M_\odot \le 0.8$.
\section{Star formation scenario in W5 E}
The W5 E region is an excellent example of triggered star formation. The presence of central O7 MS star
HD 18326 and several bright-rim like structures makes this region an interesting object to study triggered star
formation. This region contains two of the BRCs catalogued in SFO 91.
The millimeter line study by Niwa et al. (2009) has revealed the presence of a $^{13}$CO cloud
and C$^{18}$O cores on the eastern and northern sides of the region but there is no evidence for
clouds on the southern side. They also found high density
clumps in the BRC NW region. Karr \& Martin (2003) concluded that the exciting 
O stars and YSOs along the edges of the whole W5 E H{\sc ii} region
belong to two different generations. Based on the expansion velocity of the H{\sc ii} region and the
evolutionary stages of the IRAS YSOs, they concluded that the timescale is consistent with the triggering
by the RDI process.

In our earlier work (Paper I and Paper II) 
we have studied the star formation scenario in and around BRCs using the YSOs selected on the 
basis of H$\alpha$ emission and NIR excess and found evidence for triggered star 
formation around BRCs 13 and 14. Nakano et al. (2008) detected several H$\alpha$ 
emission stars in the W5 E region. They found that the young stars near the exciting 
stars are systematically older ($\sim$4 Myr) than those near the edge of the H{\sc ii} 
region ($\sim$1 Myr) and concluded that the formation of stars proceeded sequentially from 
the centre of the H{\sc ii} region to the eastern bright rims. Based on the observations of 
IR excess stars, we proposed in Paper II the occurrence of a series of RDI events in this particular 
region.

Koenig et al. (2008) have carried out an extensive $Spitzer$ survey of the whole IC 1848 region. 
On the basis of 
the large-scale distribution of YSOs selected using $Spitzer$ data, they found that 
the ratio of Class II to Class I sources within the H{\sc ii} region cavity
is $\sim$ 7 times higher compared to the regions associated with the molecular cloud. 
They attributed this difference to an age difference between the two locations and concluded 
that there exist at least two distinct generations of stars in the region. They stated that the 
triggering is a plausible mechanism to explain the multiple events of star formation and 
suggested that the W5 H{\sc ii} region merits further investigation. 

Earlier works studying the star formation scenario in the region were often qualitative 
(e.g., Karr \& Martin 2003, Koenig et al. 2008). Some were quantitative studies, but they were based on a 
smaller sample of stars (e.g., Nakano et al. 2008, Papers I and II). 
Since in the present work we have a larger database, it will be worthwhile studying 
the global scenario of star formation in the region. In the following subsections we shall 
study the star formation scenario in the cluster as well as around the BRCs.

\subsection{Cluster region}
The CMD of the identified YSOs (Fig. 9a) and the statistically cleaned CMD (Fig. 11) reveal 
a non-coeval star formation in the cluster region. The age distribution is shown in Fig. 14. 
The mean age of the YSOs is estimated to be 1.26 $\pm$ 0.69 Myr. 
This is smaller than the average age of the YSOs obtained by Nakano et al. (2008, age $\sim$4 Myr). Nakano et al. (2008) have used only H$\alpha$ emission stars 
while the present sample is larger, including NIR and MIR excess stars in addition. 
To estimate the age of the H$\alpha$ emission stars, Nakano et 
al. (2008) used PMS isochrones by Palla \& Stahler (1999), whereas we have 
used those by Siess et al. (2000). The use of different PMS evolutionary models can introduce a 
systematic shift in the age estimation. 
 
\subsection{BRC regions}
BRCs are considered to be a sort of remnant clouds originating from the dense parts (cores) in an 
inhomogeneous molecular cloud. So, if the original core was big enough, the resultant 
BRC could have undergone a series of RDI events (Kessel-Deynet \& Burkert 2003), leaving an elongated distribution 
of young stars. The distribution and evolutionary stages of such YSOs could 
be used 
to probe the star formation history in the H{\sc ii} regions. Using $Spitzer$ observations 
Koenig et al. (2008) have identified clustered and distributed population of YSOs in W5 (see their 
figure 12). The clustered population shows a nice alignments of Class II sources towards the 
directions of BRCs from the ionization sources. In Paper II, we also examined the global 
distribution of YSOs, 
the radial distribution of $A_V$ as well as the amount of NIR excess $\Delta$$∆(H-K)$, defined as the 
horizontal displacement from the middle reddening vector in the NIR-CC diagram (see Fig. 6), 
in the BRCs 13 and 14 regions. On the basis of these distributions, star formation was 
found to have propagated from the ionising source in the direction of the BRCs.  

Fig. 15 shows the global distribution of clustered YSOs selected from the study of Koenig et al. 
(2008), in the whole W5 E region, which indicates an 
alignment of YSOs towards the direction of the BRCs. Fig. 16 shows 
the radial 
variation of $\Delta$$(H - K)$ and $A_V$, of the YSOs located within the strip towards 
the direction of BRC NW region.  The distribution reveals higher $A_V$ values near the BRC NW region.

On the basis of the global distribution of YSOs in the region and the radial distribution of the amount 
of NIR excess $\Delta$$(H - K)$ and $A_V$ in the region (cf. Figs. 16, 17 and figures 6, 7 and 8 of Paper II) as well as the age distribution of YSOs, it seems that a series of RDI processes proceeded in the past from near 
the central O star towards the present locations of the BRCs.

\subsubsection{\bf Small scale sequential star formation ($S^4F$)}
\begin{table}
\caption{Mean age inside/on the rim and outside rim}
\label{age_in_out}
\begin{tabular}{ccc} \hline
Region &Mean age (Myr)    \\ 
  \hline
     & Inside/On the rim (number of sources)& Outside rim (number of sources)\\
\hline
BRC NW  &0.92 $\pm$ 0.56 (12)& 1.29 $\pm$ 0.54 (19)\\
BRC 13  &1.61 $\pm$ 1.41 (10)& 2.44 $\pm$ 1.37 (24)\\
BRC 14  &1.01 $\pm$ 0.73 (18)& 2.32 $\pm$ 1.22 (58)\\ 
\hline
\end{tabular}
\end{table}

There has been qualitative evidence for the $S^4$F hypothesis such as an 
asymmetric distribution of 
probable TTSs (Ogura et al. 2002) and of the properties of NIR excess stars 
(Matsuyanagi et al. 2006). Papers I and II presented some quantitative evidence 
on the basis of $BVI_C$ photometry. In the present 
study  the sample of YSOs is significantly larger as compared 
to that used in Papers I and II.  To further verify the $S^4$F hypothesis 
we follow the approach as given in Papers I and II . We have divided the 
YSOs  associated with BRCs into two groups: those lying inside/on the rim (i.e., 
stars embedded or projected onto the molecular cloud/lying on the rim) and outside the rim 
(i.e., stars lying outside the rim in the  projected H{\sc ii} region) (see e.g., 
Figure A1 of Paper II for BRCs 13 and 14). Mean age 
has been calculated for these regions. Since the ionising source of the BRCs studied here 
has a maximum MS lifetime of 4-5 Myr, 
therefore the sources having age greater than 5 Myr can not be expected 
as the products of triggered star formation. They might  have formed spontaneously 
in the original molecular cloud prior to the formation of the H{\sc ii} region. 
Some of them may be background stars; larger distances make them look older 
in the CM diagram. Therefore, while calculating the mean age we have not 
included those stars. The results are given in Table 5, which shows that 
in  all the BRCs, the YSOs lying on/inside the rim are younger than those 
located outside it. The above results are exactly the same as those 
obtained in Papers I and II. Therefore, the present results further 
confirm the $S^4$F hypothesis. As in Papers I and II, we again find 
a large scatter in the stellar age  in spite of a clear trend of 
the mean age. Possible reasons for the scatter, as discussed in earlier 
papers, include photometric errors, errors in extinction correction, light 
variation of young stars, drift of stars due to their proper motions, binarity of the stars, etc. 
Photometric errors and light variation as big as 0.5 mag would affect 
stellar ages by 0.25 dex, so they do not seem to be the major reason 
for the scatter. As for the extinction correction, it probably does not 
affect the results much, since in the $V$, ${(V −- I_c)}$ CMD  
the reddening vector is nearly parallel to the isochrones. As discussed 
in Papers I and II, we speculate that the proper motions of the newly born stars 
may probably be the main cause of the scatter.

\section{ EVOLUTION OF DISKS OF T-TAURI STARS}

H$\alpha$  emission and IR excess are important signatures of young PMS stars. 
These signatures in CTTSs indicate the existence of a well-developed 
circumstellar disk actively interacting with the central star. Strong 
H$\alpha$  emission (equivalent width EW $>$ 10 \AA) in CTTSs is attributed 
to the magnetospheric accretion of the innermost disk matter onto the 
central star (Hartmann et al. 1994; Edwards et al. 1994; Muzerolle 
et al 2001 and references therein). On the other hand the weak H$\alpha$  
emission (EW $<$ 10 \AA) in weak-line T Tauri stars (WTTSs), which lack disks 
(or, at least inner disks), is believed to originate from their 
chromospheric activity (e.g., Walter et al. 1988; Mart{\'{\i}}n 1998). In the 1990s, 
a large number of WTTSs were found in and over wide areas around 
T-associations by X-ray surveys with ROSAT, which led to active 
studies on the nature of the so-called dispersed WTTSs. 

The ``standard model'' (Kenyon and Hartmann 1995) postulates that the 
CTTSs  evolve to the WTTSs by losing their circumstellar disk (or, 
at least its inner part). The age distribution derived from the HR 
diagram of the Taurus region indicates that the WTTSs are 
systematically older than the CTTSs, but the statistical significance 
is low (Kenyon and Hartmann 1995; Hartmann 2001; Armitage et al. 2003). 
Bertout et al. (2007) concluded that the observed age and mass distribution of CTTSs and 
WTTSs in the Taurus-Auriga T association can be explained by assuming that a CTTS evolves 
into a WTTS. In Paper II we compared the age distribution of CTTSs and WTTSs 
associated with several BRCs and supported the conclusion of Bertout 
et al. (2007).

On the other hand, there have also been many observations which 
claimed that the CTTS and the WTTS are coeval and have indistinguishable 
age distribution (e.g., Walter et al. 1988; Lawson et al. 1996; 
Gras-Vel{\'a}zquez \& Ray 2005). From the analyses of the HR diagram 
of the CTTSs and WTTSs in Chamaeleon I, Lawson et al. (1996) concluded 
that some stars may be born even almost diskless
or lose the disk at very early stages (age $<$ 1 Myr). 

In the present work we have derived the age of 79, 26, 29, 59 Class II and 27, 5, 5, 17 
Class III sources in the cluster, BRC NW, BRC 13, BRC 14 regions, 
respectively. Assuming that all the identified Class II and Class III stars 
are CTTSs and WTTSs respectively, we can study the possible evolution of the TTSs. 
The advantage of our sample in addressing this issue is that the stars are spatially, i.e., 
three-dimensionally, very close to each other, so there should be no problem of distance difference, 
contrary to extended T associations. Here we would like to caution the readers that although we 
have attempted to clean the Class III sample from sources not belonging to W5, however, to confirm the WTTSs  nature of these sources in the absence of spectral information such as measurement of the Lithium absorption line is difficult.

Since data for Class III sources in the BRC NW and BRC 13 regions are not statistically significant, in  Fig. 17  we show  the cumulative distribution of Class II and Class III sources in the cluster and BRC 14 region only. 
Fig. 17 indicates that both the Class III and Class II sources have rather similar distribution.  A Kolmogorov-Smirnov (KS) test
for the combined sample of  the cluster and BRC 14  region
confirms the statement that the cumulative distributions of Class II sources (CTTSs) and Class III sources (WTTSs)  are different only at a confidence level of 50\%.
Hence, we can conclude that both the samples have rather similar distribution with practically indistinguishable age distribution.
This result is  in contradiction with that of Bertout et al. (2007) for the Taurus-Auriga 
T-association and that of Paper II, that WTTSs are older than CTTSs and CTTSs evolve into WTTSs; and it is in agreement with those which claim that the CTTSs and WTTSs 
are coeval and have indistinguishable properties. 
Here we have to keep in mind that the classification of CTTS and WTTS in the present work is 
based on the Spitzer MIR observations whereas in Paper II and in Bertout et al. (2007), 
the classification was based on the EWs of H$\alpha$ emission stars. H$\alpha$  surveys may fail 
to detect Class III sources which have smaller EWs whereas, those sources can be identified using the 
MIR observations. Here it is worthwhile to point out that the Class III sources  by Koenig et al. 
(2008)  were identified on the basis of all four $Spitzer$ IRAC bands (3.6, 4.5, 5.8 and 8.0 $\mu$m). 
The detection of Class III sources would be suppressed  because of less sensitivity at  5.8 and 8.0 $\mu$m 
bands and also because of the very bright, variable background at these wavelengths in W5. 
Koenig et al. (2008) have stated that their photospheric sample is only 90\% complete to 
2 M$_\odot$ in general in W5, and to 8 M$_\odot$ on regions of bright background emission 
(e.g., the BRCs studied in this paper). This fact is evident in  $V/ (V-I)$ CMD (Fig. 7)  by 
comparing the number of Class II sources to the number of Class IIIs fainter than 20th magnitude in $V$  band. 

\section{Conclusions}
In this paper we have made $VI_c$ photometric studies of the 
newly identified central cluster of W5 E. NIR data from 2MASS and MIR data from IRAC/MIPS of 
the $Spitzer$ telescope have also been added to consider the 
properties of the cluster. Also, by incorporating $Spitzer$ data 
we have re-analysed the properties (age and spatial distribution, etc.) 
of the YSOs in W5 E with special emphasis on the three BRC regions. We 
obtained the following as the main conclusion of the present work. 

The central cluster has a distance of $\sim$2.1 kpc, a radius of $6^\prime$ and 
a mean age of $\sim$ 1.3 Myr. The reddening law is normal. The star formation in the 
cluster region is found to be non-coeval, with an age spread of $\sim$ 
5 Myr. The slope of the IMF within the cluster region in the mass range $0.4 \le M/M_\odot \le 30$ 
is found to be $\Gamma = -1.29 \pm 0.03$, which is comparable to the Salpeter value. In the mass range of $0.2 \le M/M_\odot \le 0.8$; the CMFs of the YSOs associated with the three BRCs 
show a break in the slope at $\sim$ $0.8$ $M_\odot$. The CMFs indicate that the  cluster region 
and BRC NW have relatively more low mass YSOs in the mass range $0.2 \le M/M_\odot \le 0.8$.

The distribution of the YSOs in the W5 E region indicates that they are globally aligned from the 
ionising source towards the BRCs. A comparison of the average age of the YSOs lying on/inside 
and outside the bright rim indicates a quantitative age gradient in the sense that the  YSOs 
located on/inside the rim are younger than those located outside the rim. The results are 
similar to those reported in Paper I and Paper II. These results confirm the $S^4$F 
hypothesis. The globally aligned distribution of the YSOs, indicate that a series of RDI 
events took place in the past from near the central ionising source to 
the periphery of the H{\sc ii} region.

In the present study it is  found that the age distributions of  the Class II  and 
Class III sources are the same.  This result is in accordance with the results which 
claimed that the CTTS and the WTTS are coeval and have indistinguishable age distribution 
(e.g., Walter et al. 1988; Lawson et al. 1996; Gras-Velzquez \& Ray 2005).
\section{ACKNOWLEDGMENTS} 
The authors are grateful to the anonymous referee for useful comments that improved the contents of the paper. We are thankful to the Kiso Observatory, IAO and ARIES for allotting the observing time. 
We also thank the staff of Kiso Observatory (Japan), IAO, Hanle and its remote control station 
CREST, Hosakote for their assistance during observations. This publication makes use 
of data from the Two Micron All Sky Survey, which is a joint project of  the University 
of Massachusetts and the Infrared Processing and Analysis Center/California Institute of
Technology, funded by the National Aeronautics and Space Administration and the National 
Science Foundation. NC is thankful for the financial support for this study 
through the fellowships granted by the DST and CSIR, India.

\newpage
\begin{figure*}
\centering
\includegraphics[scale = .87, trim = 5 5 5  5, clip]{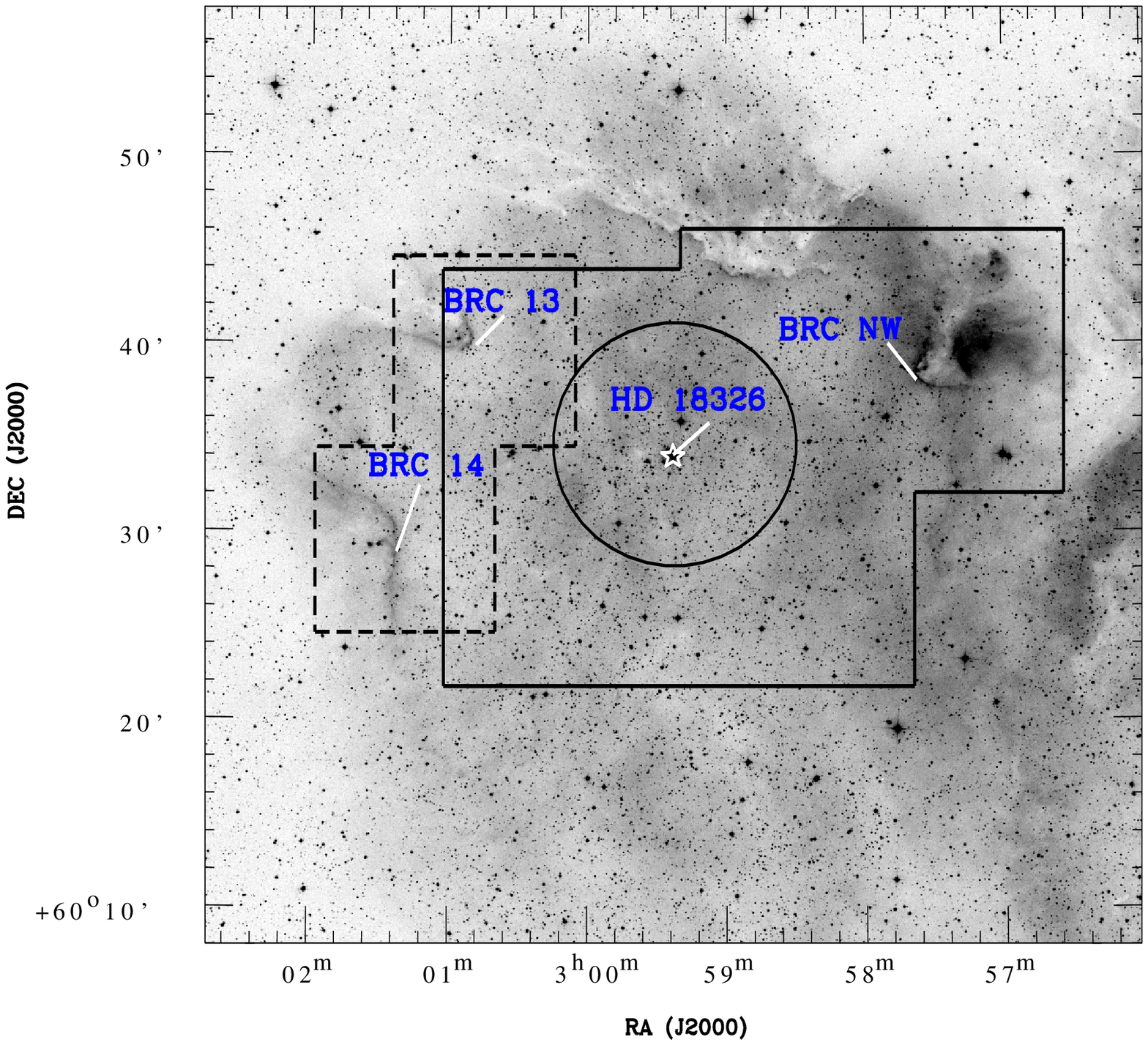}
\caption{The 50 $\times$ 50 arcmin$^2$ DSS2-R band image of the W5 E region. 
The area marked with thick lines is the region for which deep images are 
taken in $V$ and $I_c$ bands. The dashed lines represent the boundaries of the region for which we have used 
data from our earlier work (Paper II). The circle represents the boundary of the cluster. The abscissa and the ordinates are RA and DEC for the J2000 epoch.}
\label{fig1}
\end{figure*}

\begin{figure*}
\centering
\includegraphics[scale = .5, trim = 5 5 5  5, clip]{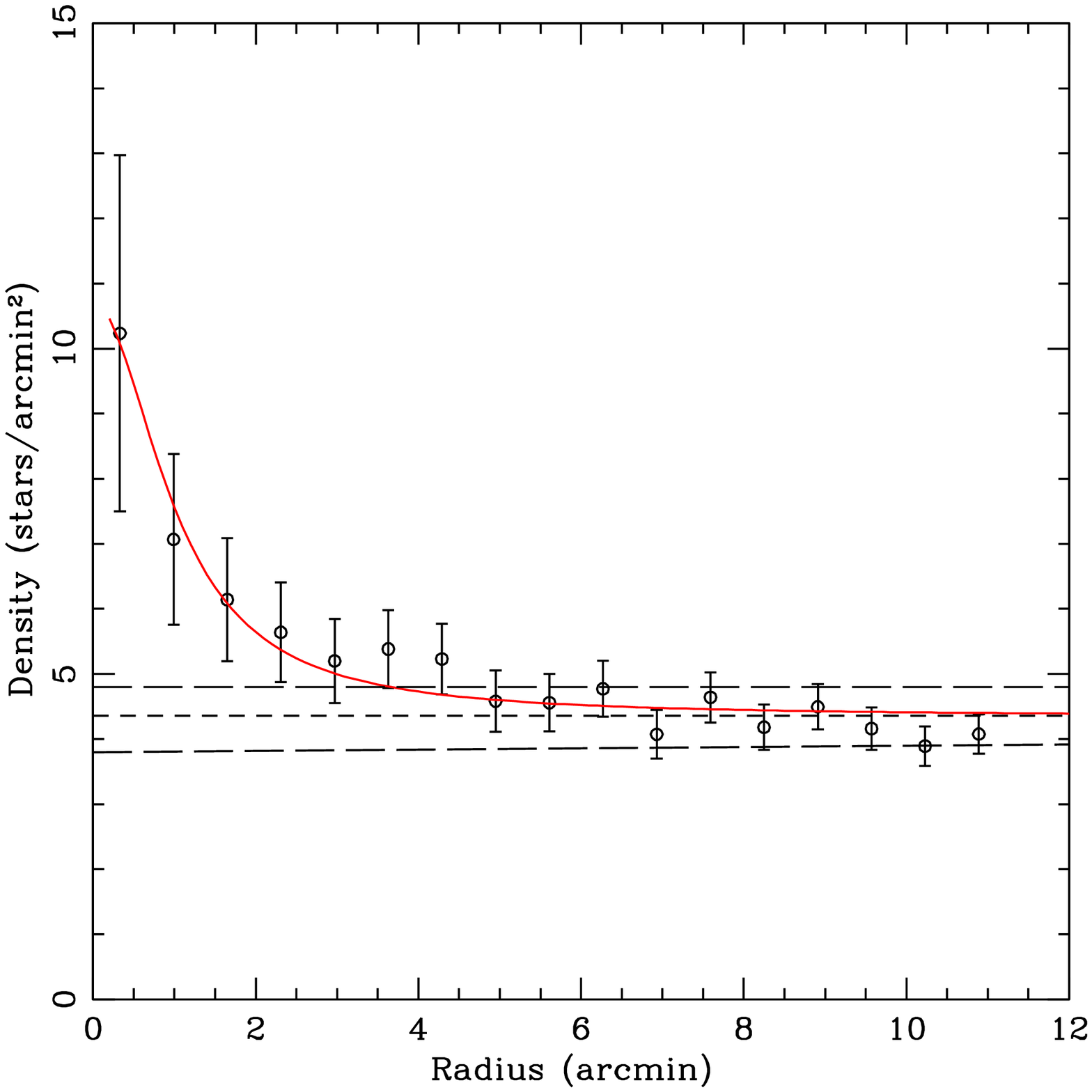}
\includegraphics[scale = .5, trim = 5 5 5  5, clip]{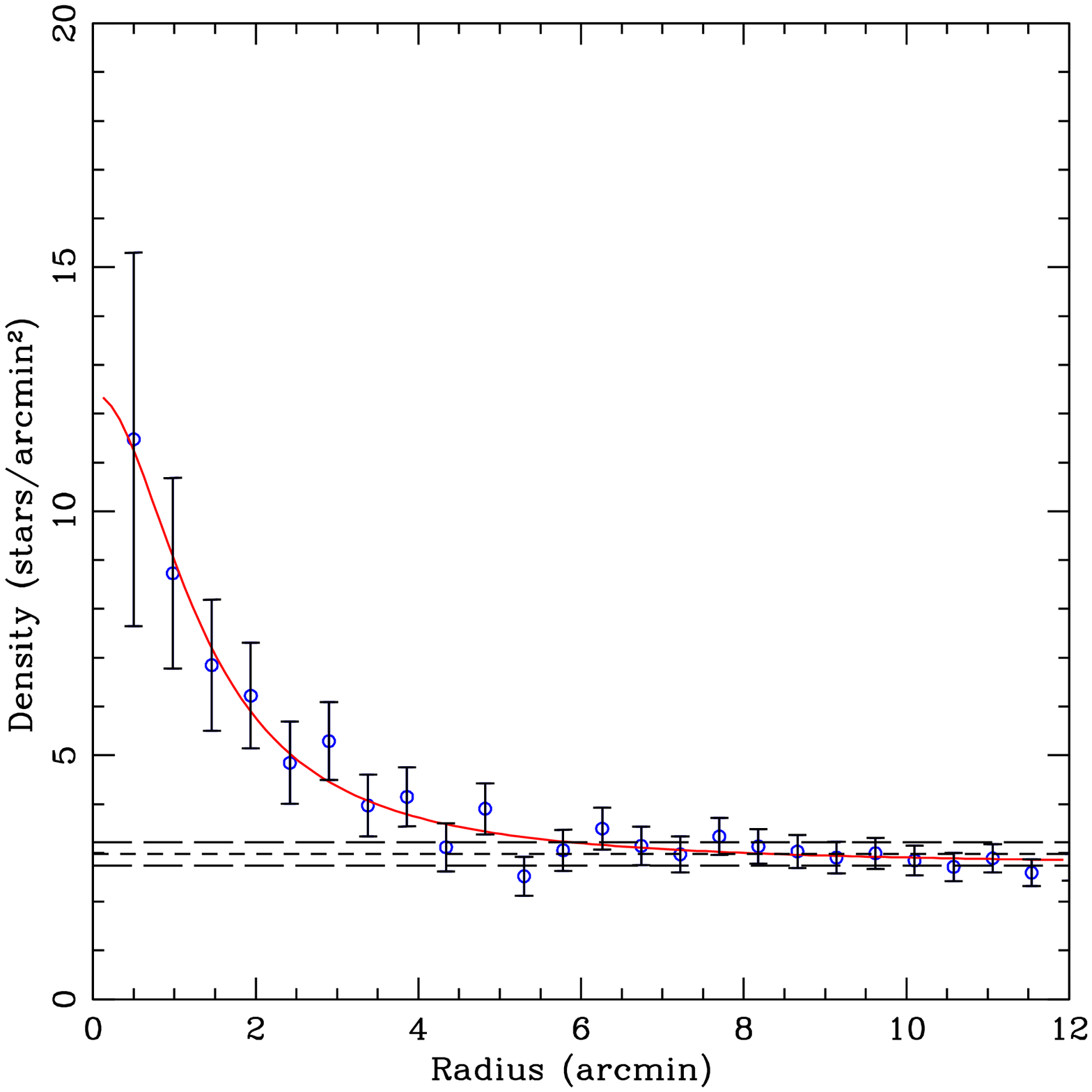}
\caption{ Radial density profiles for the cluster using the optical (upper panel) 
and 2MASS (lower panel) data. Thick dashed line represents the mean density level 
of the field stars and thin dashed lines are the error limits for the field star density. 
The continuous curve shows the least-squares fit of the King 
(1962) profile to the observed data points. The error bars represent $\pm$$\sqrt{N}$ errors.}
\label{fig2}
\end{figure*}
\begin{figure*}
\centering
\includegraphics[scale = .87, trim = 5 5 5  5, clip]{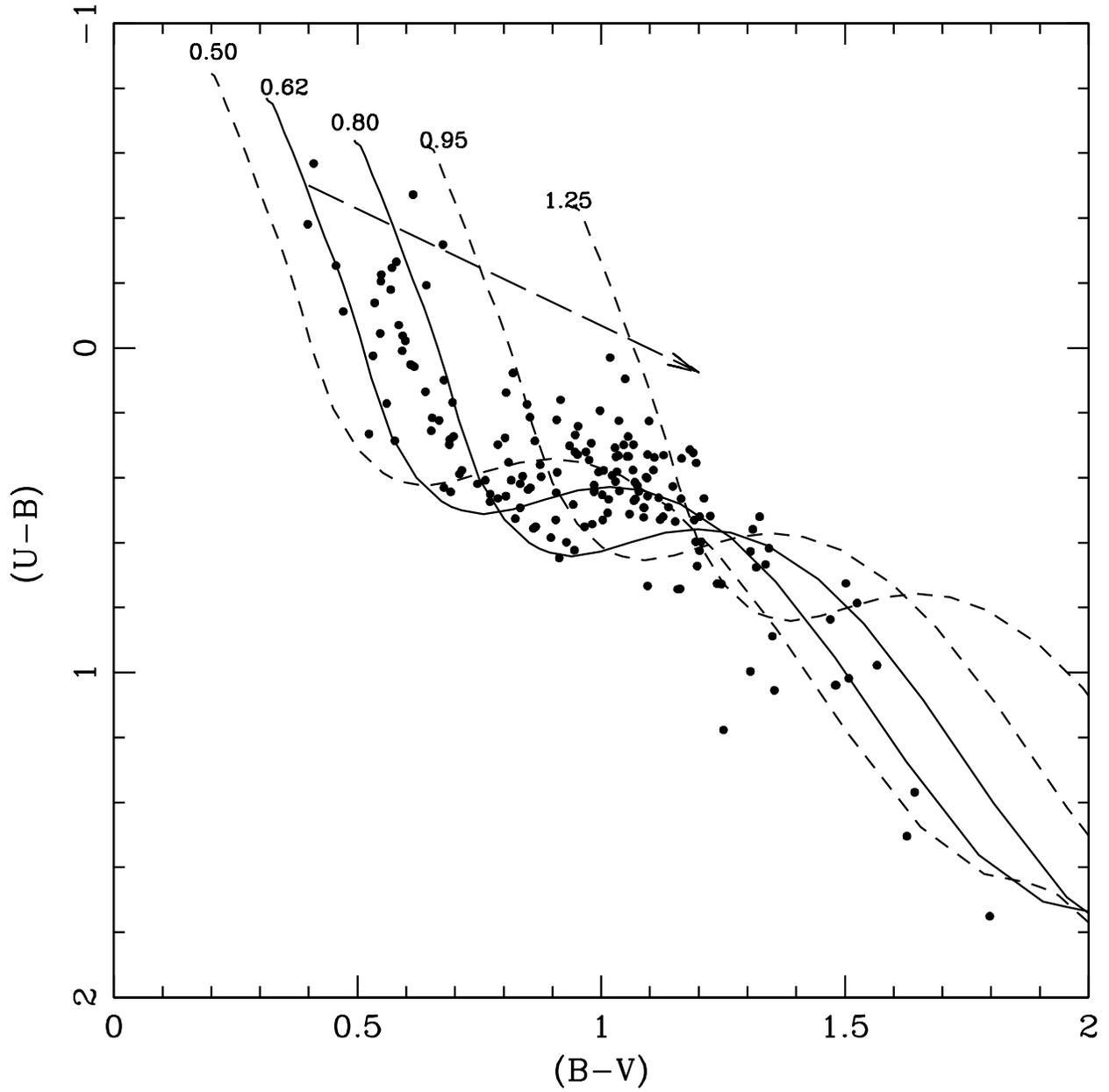}

\caption{ $(U - B)/(B - V)$ CC diagram for the stars within the cluster 
radius ($r_{cl} \le 6^\prime $).  The continuous curves represent the ZAMS by Girardi et al. (2002) 
shifted along the reddening slope of 0.72  (shown as dashed arrow) for $E(B - V )$ = 0.62 mag
and 0.80 mag, respectively. The dashed curves represent the ZAMS reddened by 
$E(B - V)$ = 0.50 mag, 0.95 mag and 1.25 mag, respectively to match the probable
foreground and background populations (see the text for details).
}
\label{fig3}
\end{figure*}
\newpage
\begin{figure}
\centering
\includegraphics[scale = .73, trim = 5 45 5 75, clip]{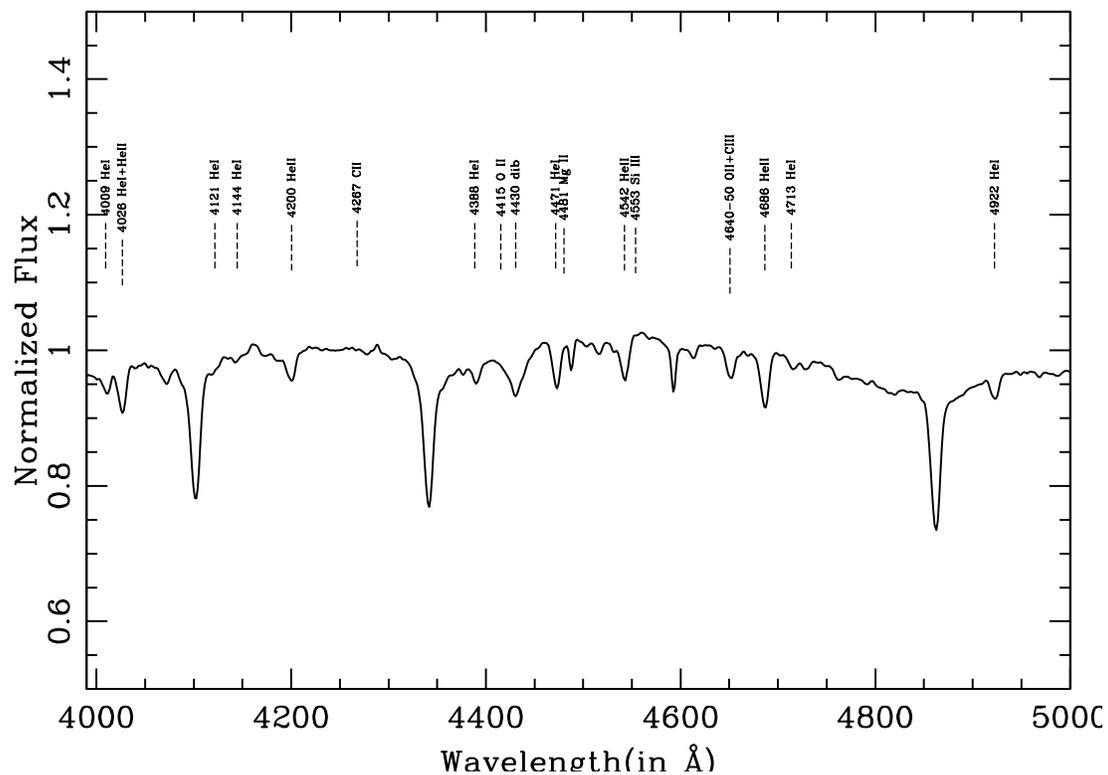}
\caption{Flux calibrated normalised spectrum of HD 18326.}
\label{fig5}
\end{figure}
\newpage
\begin{figure*}
\centering
\includegraphics[scale = .67, trim = 5 5 5  5, clip]{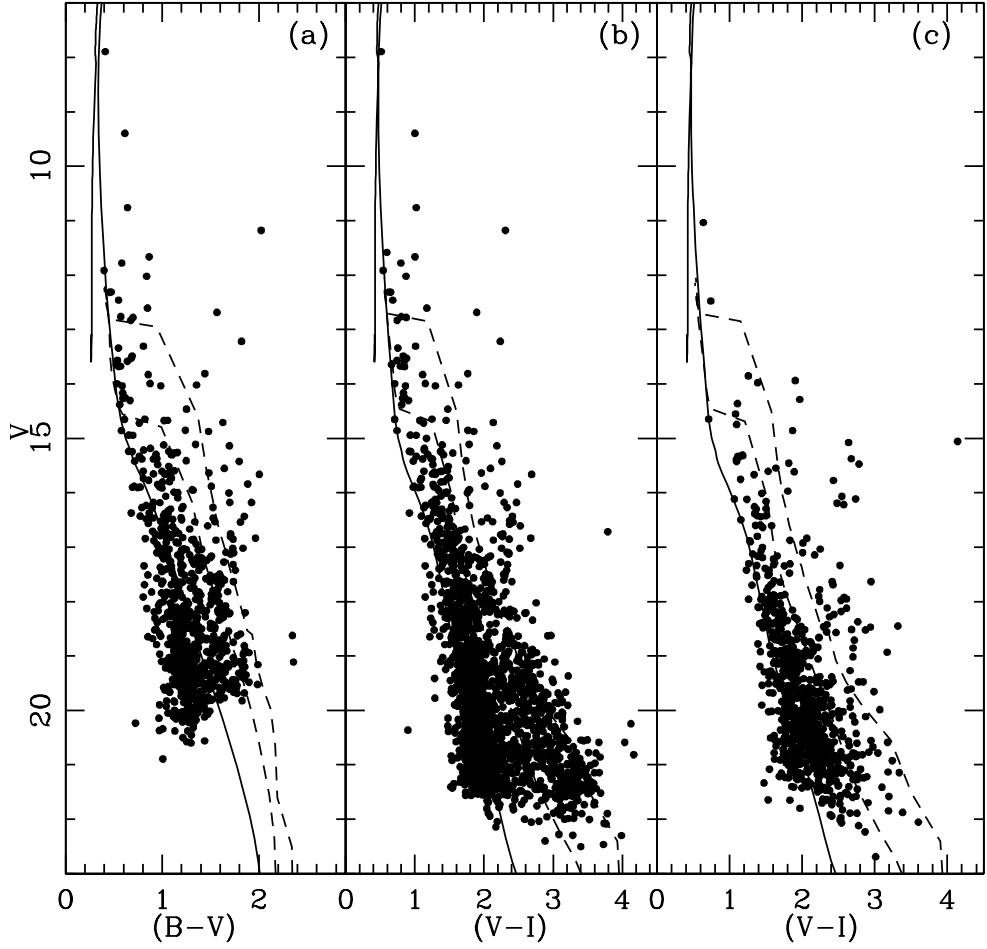}

\caption{(a) and (b): The $V/(B - V)$ and $V/(V - I)$ CMDs for the stars 
within the cluster radius; (c): $V/(V - I)$ CMD for stars in the field region 
having same area as in (a) and (b). The continuous line is the isochrone of 4 Myr from 
Girardi et al. (2002) and dashed lines are 1 Myr and 5 Myr PMS isochrones from 
Siess et al. (2000). The isochrones are corrected for the cluster distance of 2.1 kpc and reddening $E(B - V) = 0.62$ mag.}
\label{fig4}
\end{figure*}
\begin{figure*}
\centering
\includegraphics[scale = .67, trim = 5 5 5  5, clip]{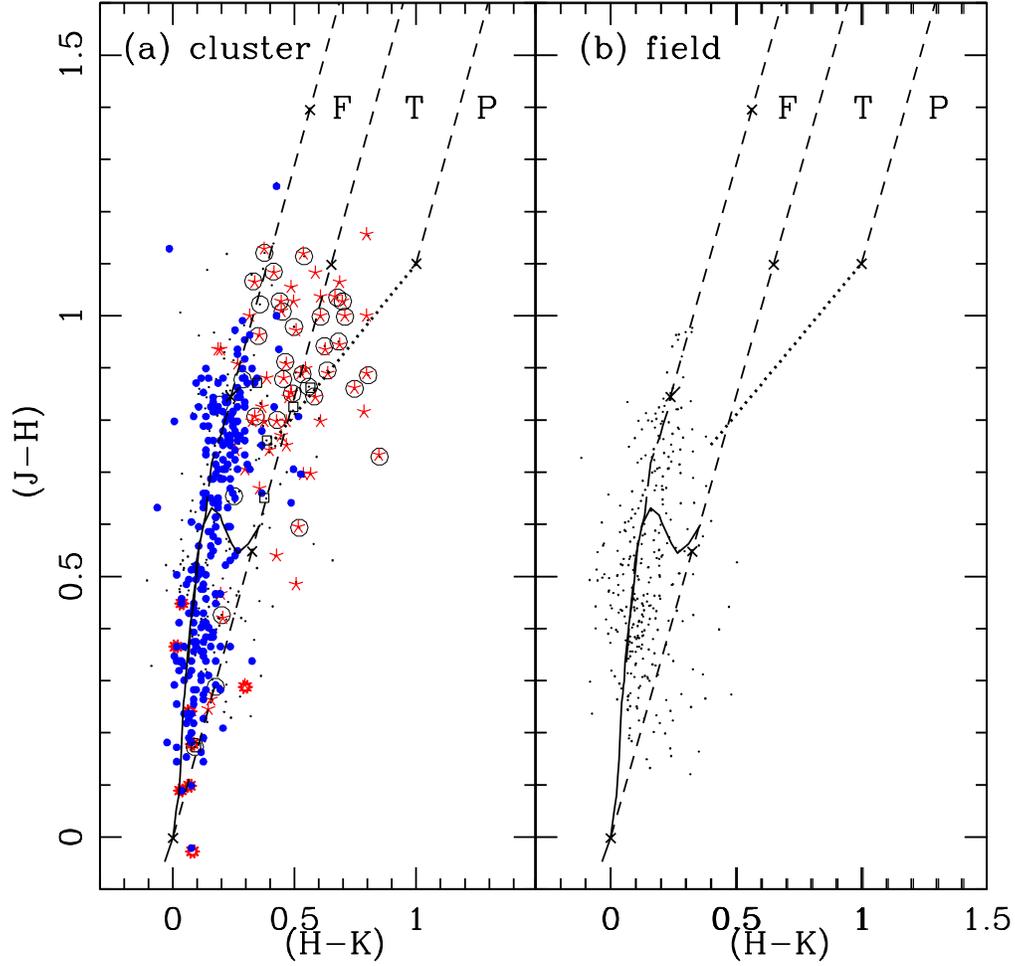}
\caption{ NIR $(J-H)/(H-K)$ CC diagrams for the stars (a) 
within the cluster radius and (b) in the reference field. Small dots represent 2MASS 
sources and open circles represent H$\alpha$ sources from Nakano et 
al. 2008. Class II, Class III and transition sources from the $Spitzer$ 
photometry are shown by asterisks, filled circles and open squares, 
respectively. Dashed straight lines represent the reddening vectors 
(Cohen et al. 1981). The crosses on the dashed lines are separated by $A_V$ = 5 mag.}
\label{fig7}
\end{figure*}
\begin{figure}
\centering
\includegraphics[scale = .67, trim = 5 5 5  5, clip]{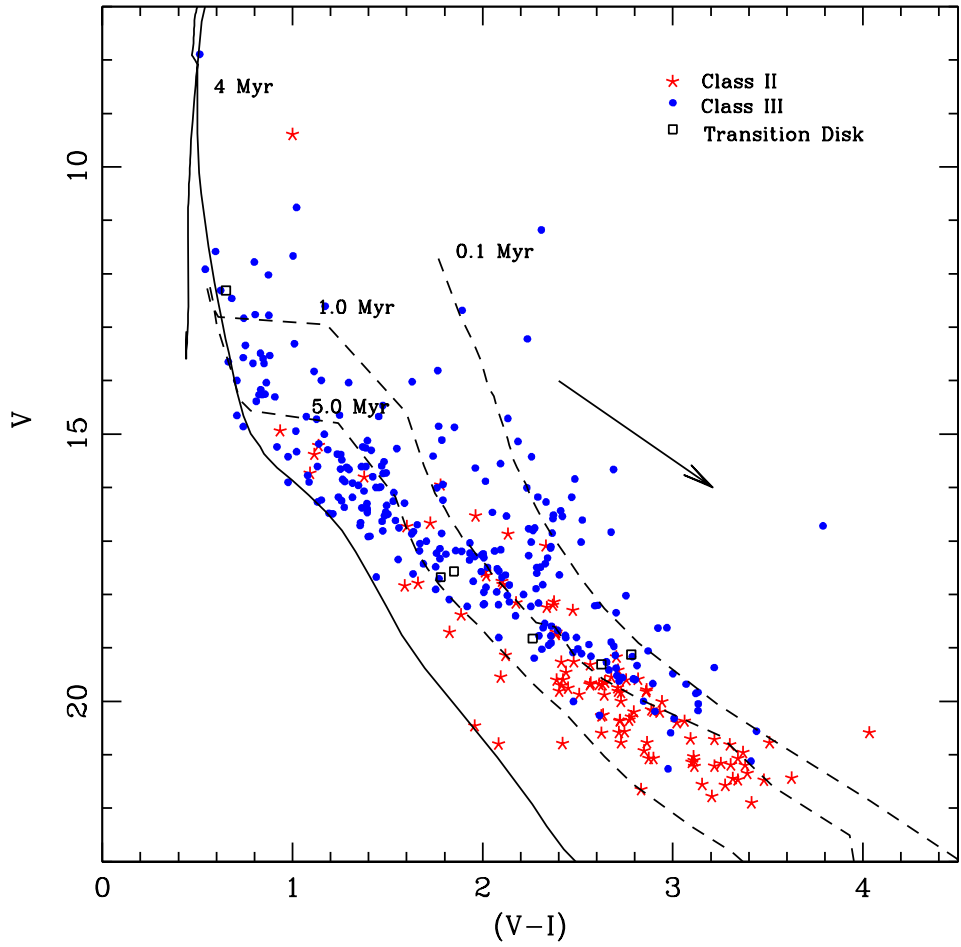}
\caption{ $V/(V - I)$ CMD for the clustered YSOs identified by Koenig et al. (2008) 
in the cluster region. The continuous line is the isochrone of 4 Myr from   
Girardi et al. (2002) and dashed lines are the 0.1 Myr, 1 Myr and 5 Myr PMS isochrones from
Siess et al. (2000). The isochrones are corrected for the cluster distance of 2.1 kpc and reddening $E(B - V) = 0.62$ mag}
\label{fig6}
\end{figure}


\begin{figure*}
\centering
\includegraphics[scale = .87, trim = 5 5 5  5, clip]{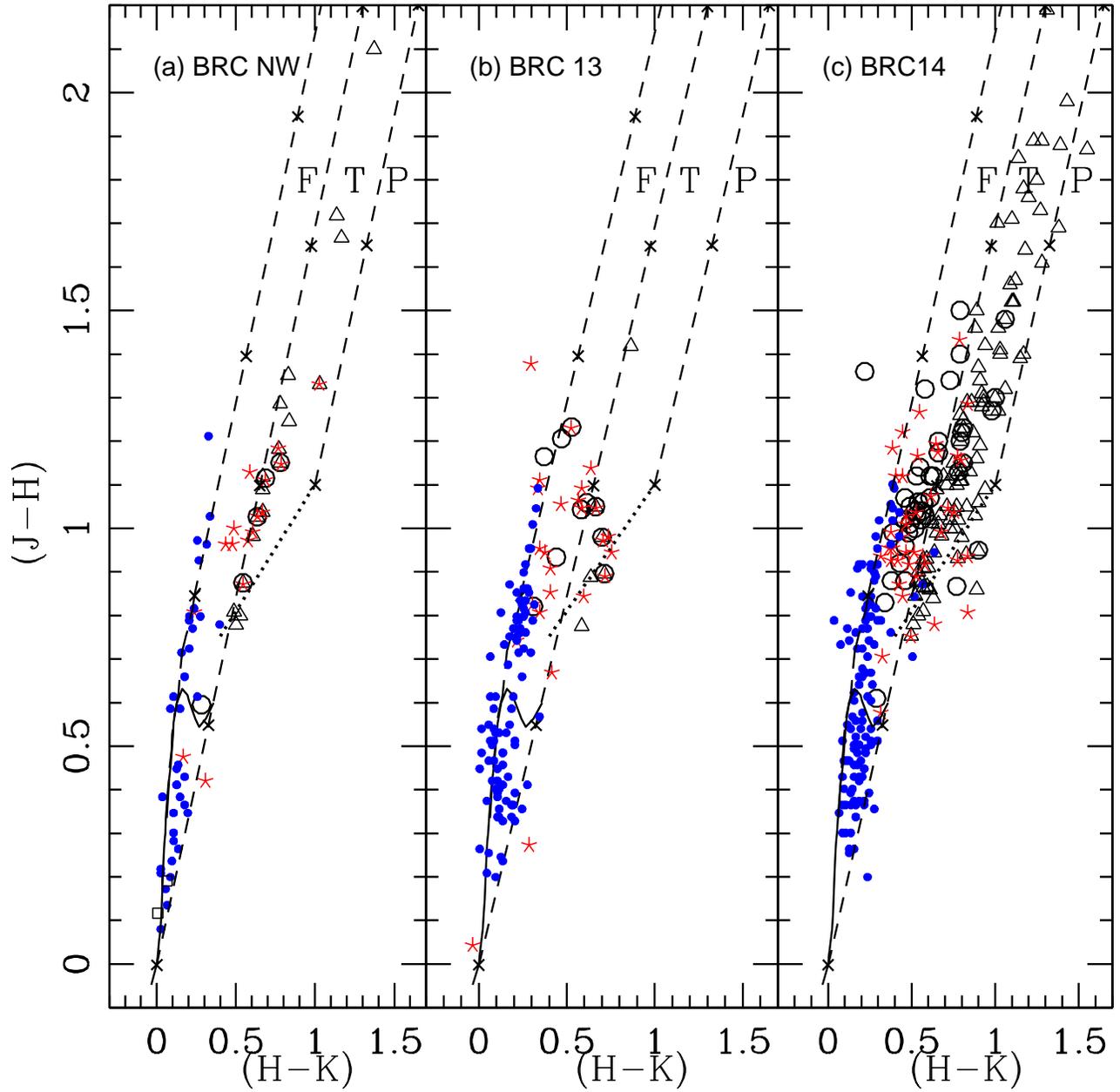}
\caption{ The $(J - H)/(H - K)$ CC diagrams of the YSOs in the BRC NW, BRC 13 and BRC 14. 
Class I, Class II, Class III and transition sources from $Spitzer$ photometry 
are shown by filled triangles, asterisks, filled circles and open squares,
respectively. NIR excess sources from 2MASS and Matsuyanagi et al. (2006) (in case of BRC 14) are shown by open triangles and 
H$\alpha$ sources from Ogura et al. (2002) are shown by open circles. }
\label{fig8}
\end{figure*}

\begin{figure*}
\centering
\includegraphics[scale = .42, trim = 15 15 15 15, clip]{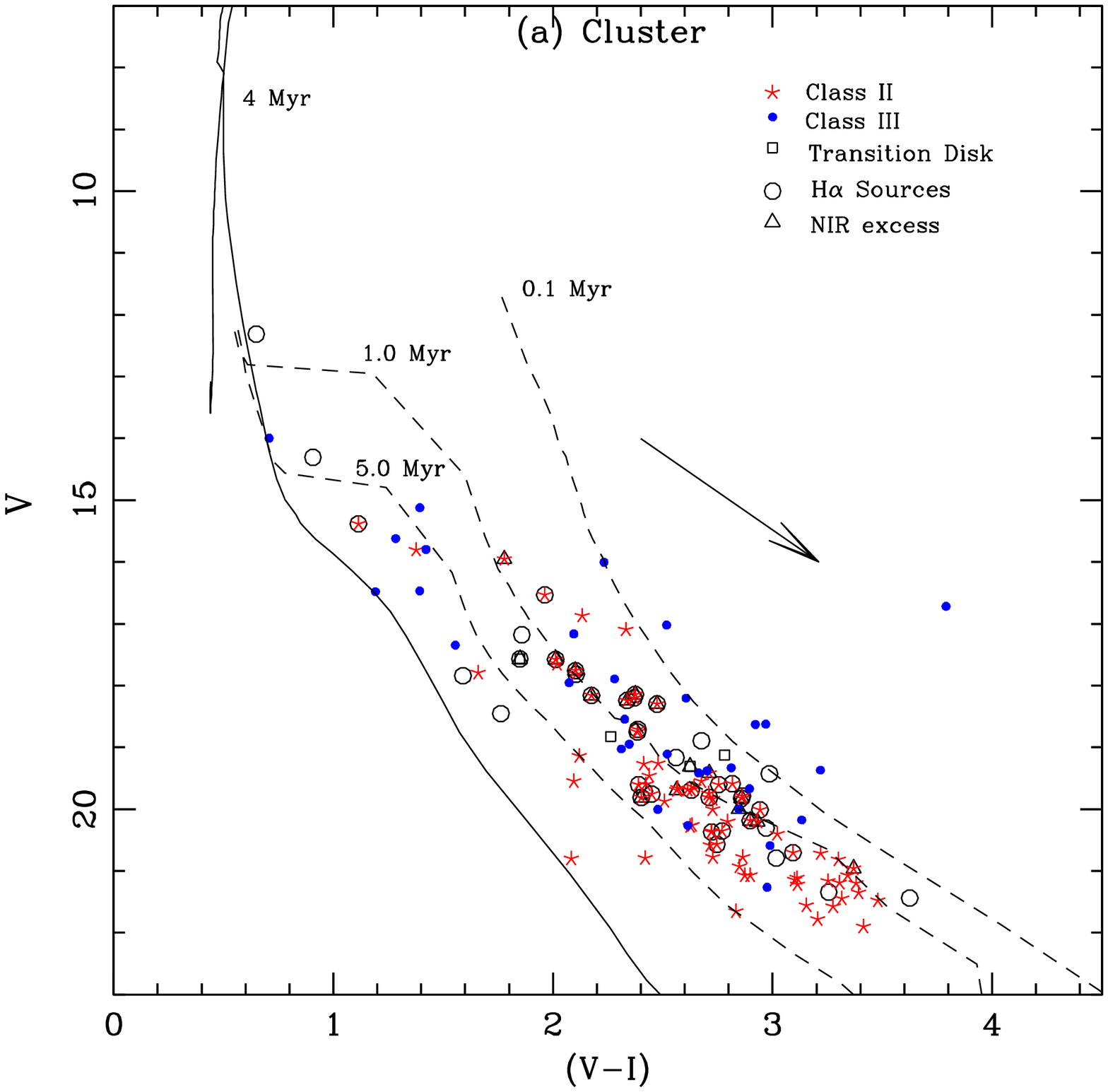}
\includegraphics[scale = .42, trim = 15 15 15  15, clip]{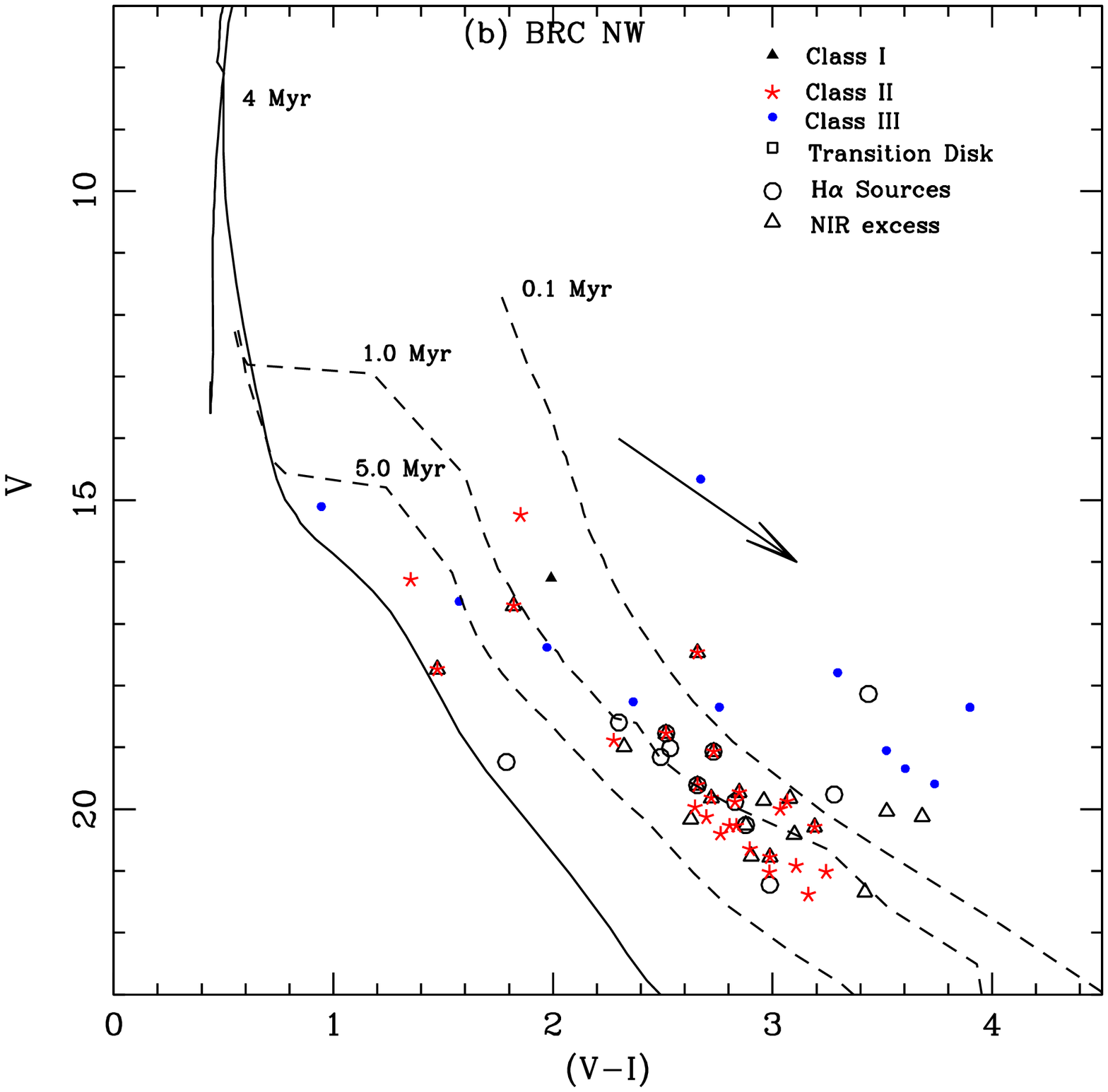}
\includegraphics[scale = .42, trim = 15 15 15  15, clip]{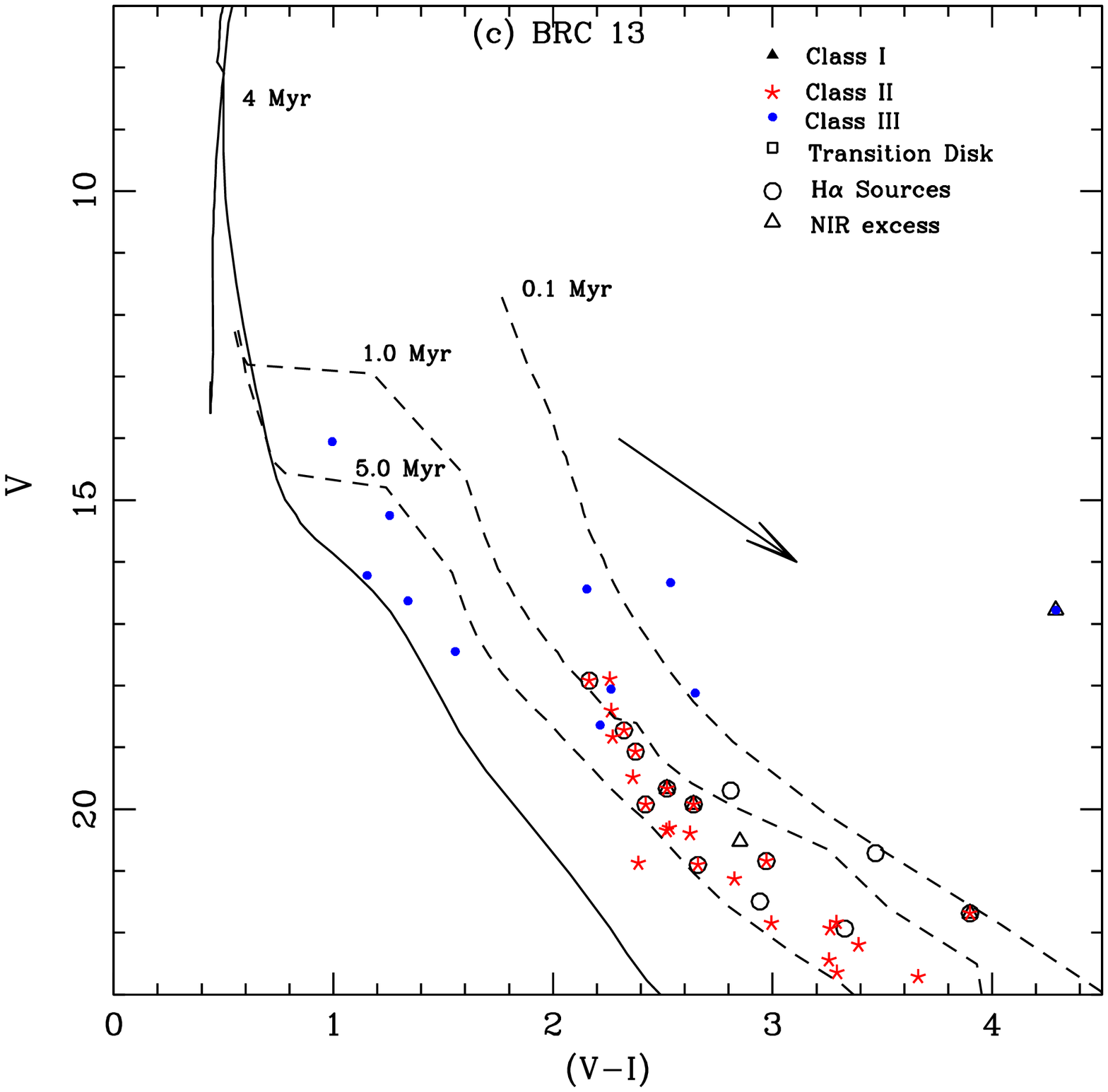}
\includegraphics[scale = .42, trim = 15 15 15  15, clip]{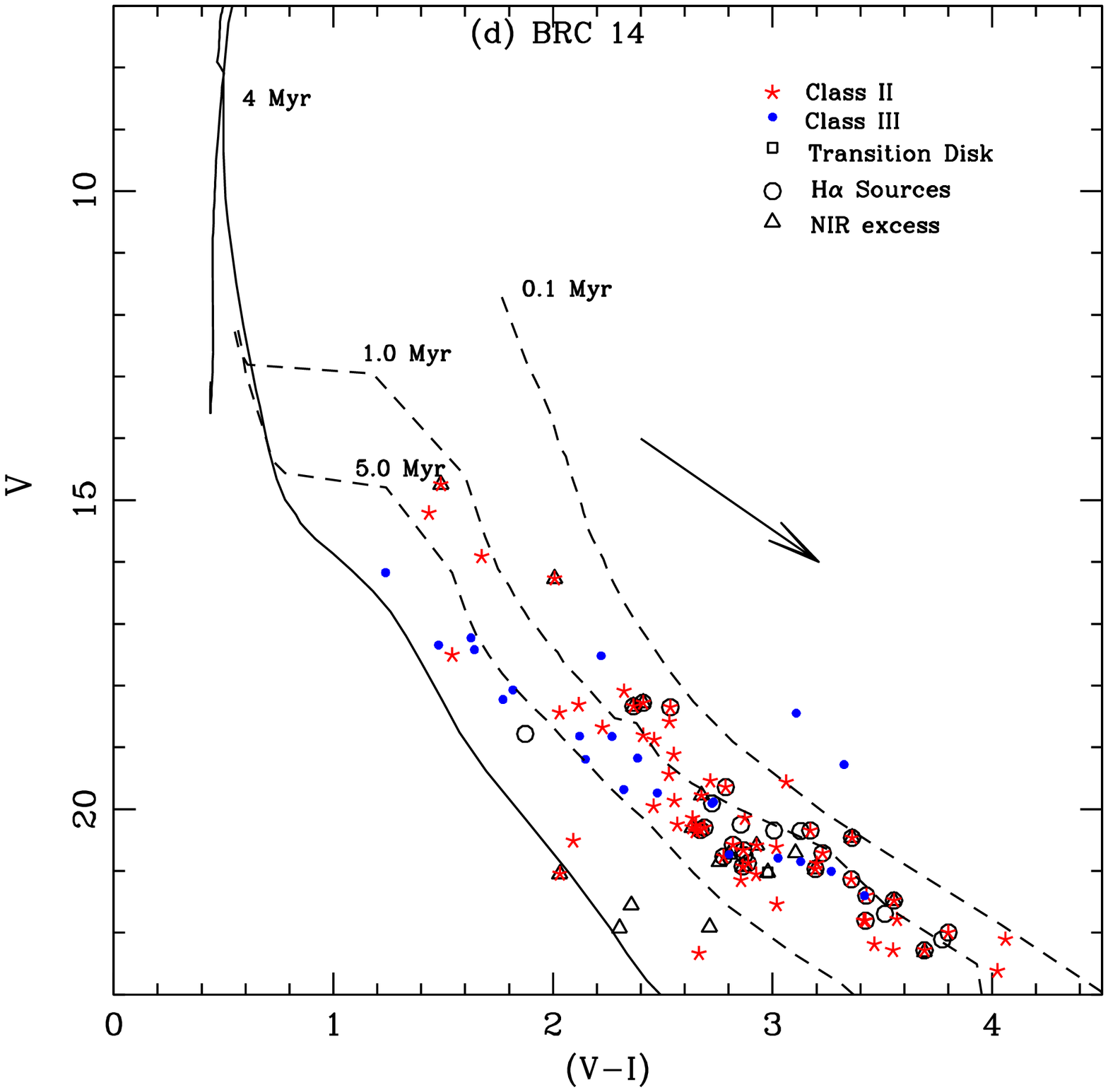}
\caption{$V/(V - I)$ colour-magnitude diagrams for the YSOs selected in the presented study (see the text) in the (a) cluster, (b) BRC NW, (c) BRC 13 and (d) BRC 14 regions. 
The continuous line is the isochrone of 4 Myr from 
Girardi et al. (2002) and dashed lines are 0.1, 1 and 5 Myr PMS isochrones from 
Siess et al. (2000). The isochrones are corrected for the distance and reddening of the respective regions. The arrow represents the reddening vector.}
\label{fig9}
\end{figure*}
\begin{figure*}
\centering
\includegraphics[scale = .76, trim = 5 5 5  5, clip]{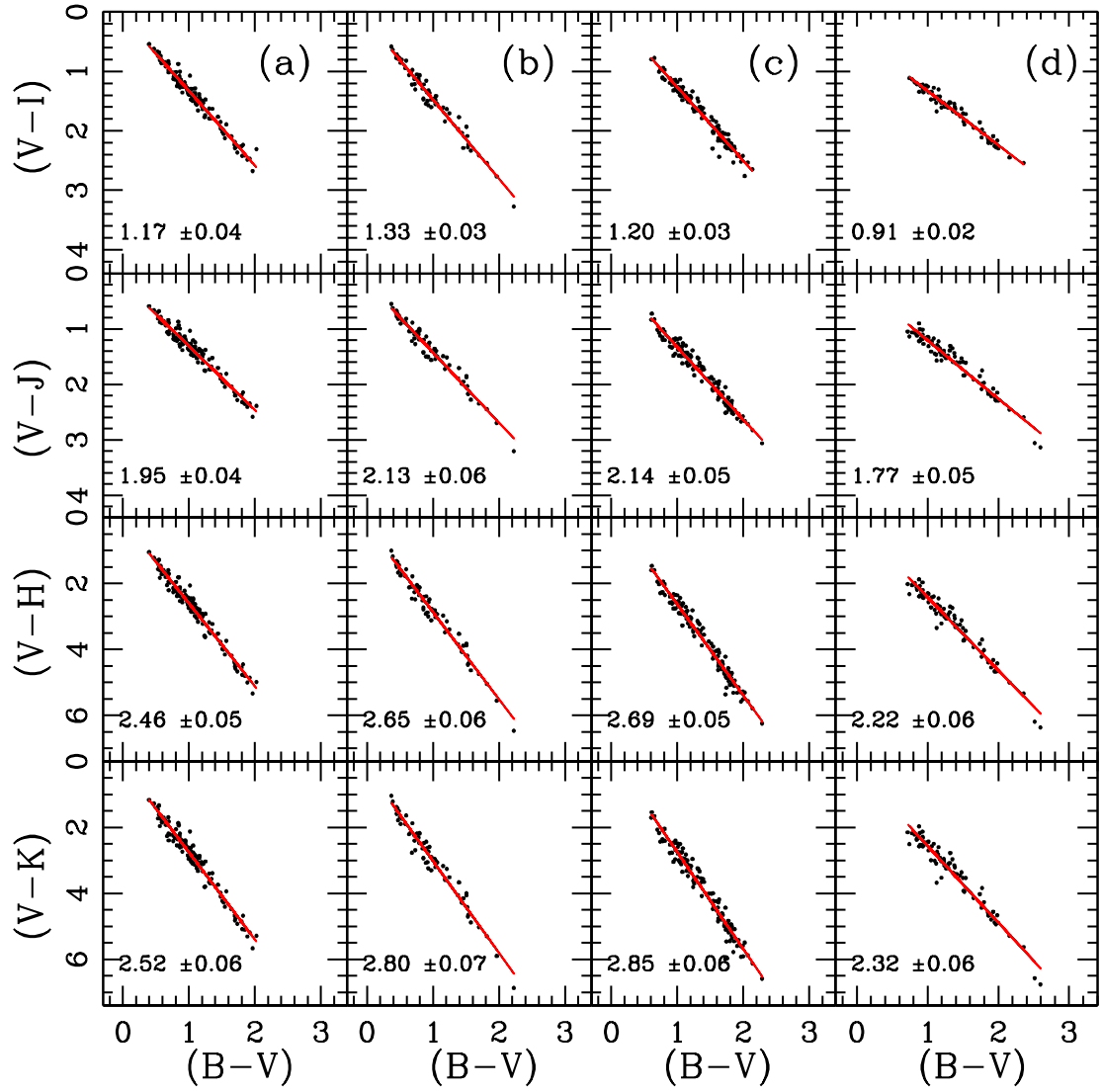}
\caption{TCDs for the diskless (Class III) sources in the (a) cluster, (b) BRC NW, 
(c) BRC 13 and (d) BRC 14  regions.}
\label{fig10}
\end{figure*}
\newpage
\begin{figure}
\centering
\includegraphics[scale = .67, trim = 5 5 5  5, clip]{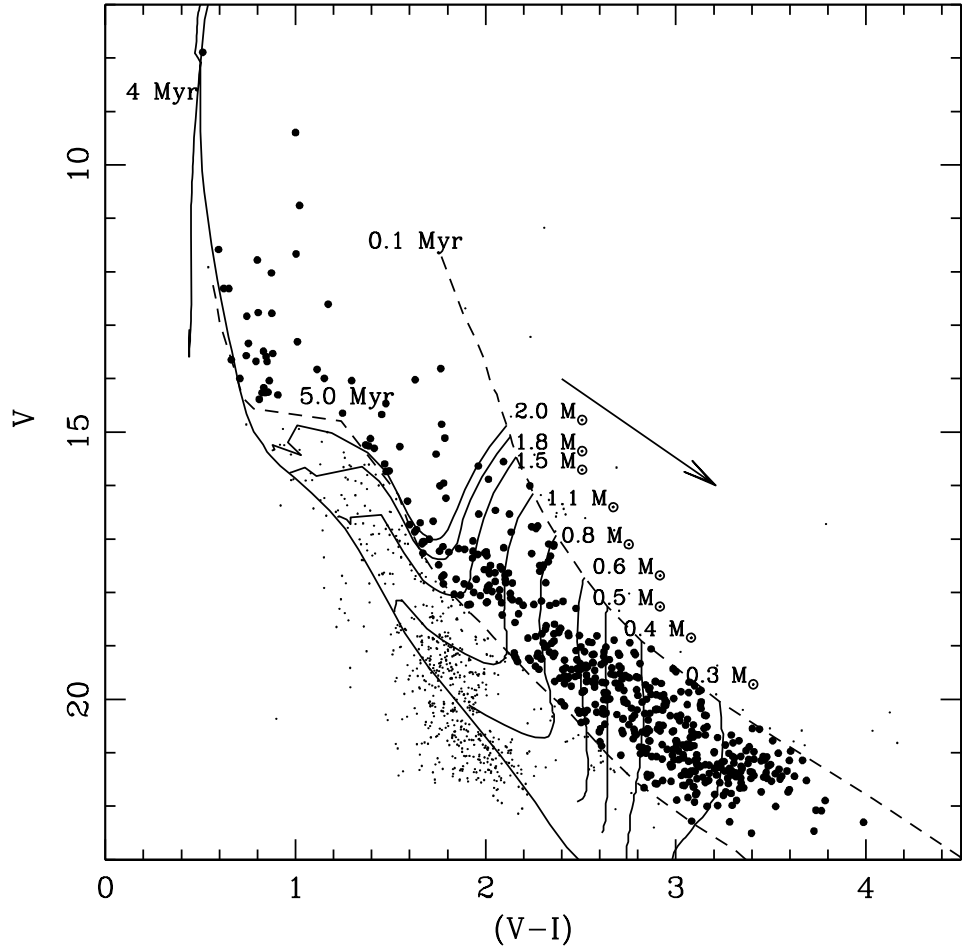}
\caption{Statistically cleaned $V /(V - I )$ CMD for stars lying within the 
cluster radius. The stars having PMS age $\le$ 5 Myr are considered as 
representing the statistics of PMS stars in the region and are shown by 
filled circles. The isochrone for 4.0 Myr age by Girardi et al. (2002) and 
PMS isochrones of 
0.1, 5.0 Myr along with evolutionary tracks for different masses by 
Siess et al. (2000) are also shown. All the isochrones are corrected for 
the cluster distance and reddening. The corresponding values of masses 
in solar mass are given at the right side of each track. Points shown
by small dots are considered as non-members. The arrow represents the reddening vector.}
\label{fig11}
\end{figure}

\begin{figure*}
\centering
\includegraphics[scale = .47, trim = 5 5 5  5, clip]{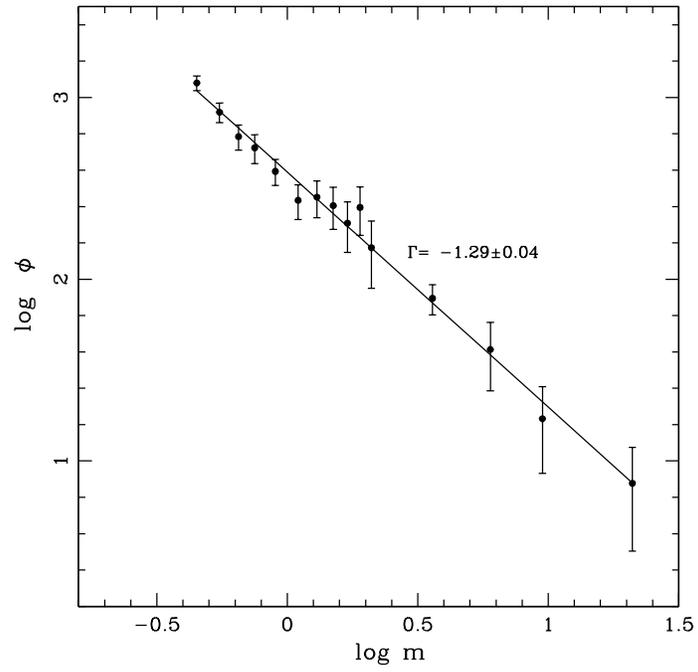}
\caption{ The MF in the cluster region derived using the optical data. 
The $\phi$ represents $N$/{\rm{d} log $m$ }. The error bars 
represent $\pm$$\sqrt{N}$ errors. The continuous line shows least-squares 
fit to the mass ranges described in the text. The value of the
slope obtained is mentioned in the figure.
}
\label{fig12}
\end{figure*}
\begin{figure*}
\centering
\includegraphics[scale = .74, trim = 5 5 5  5, clip]{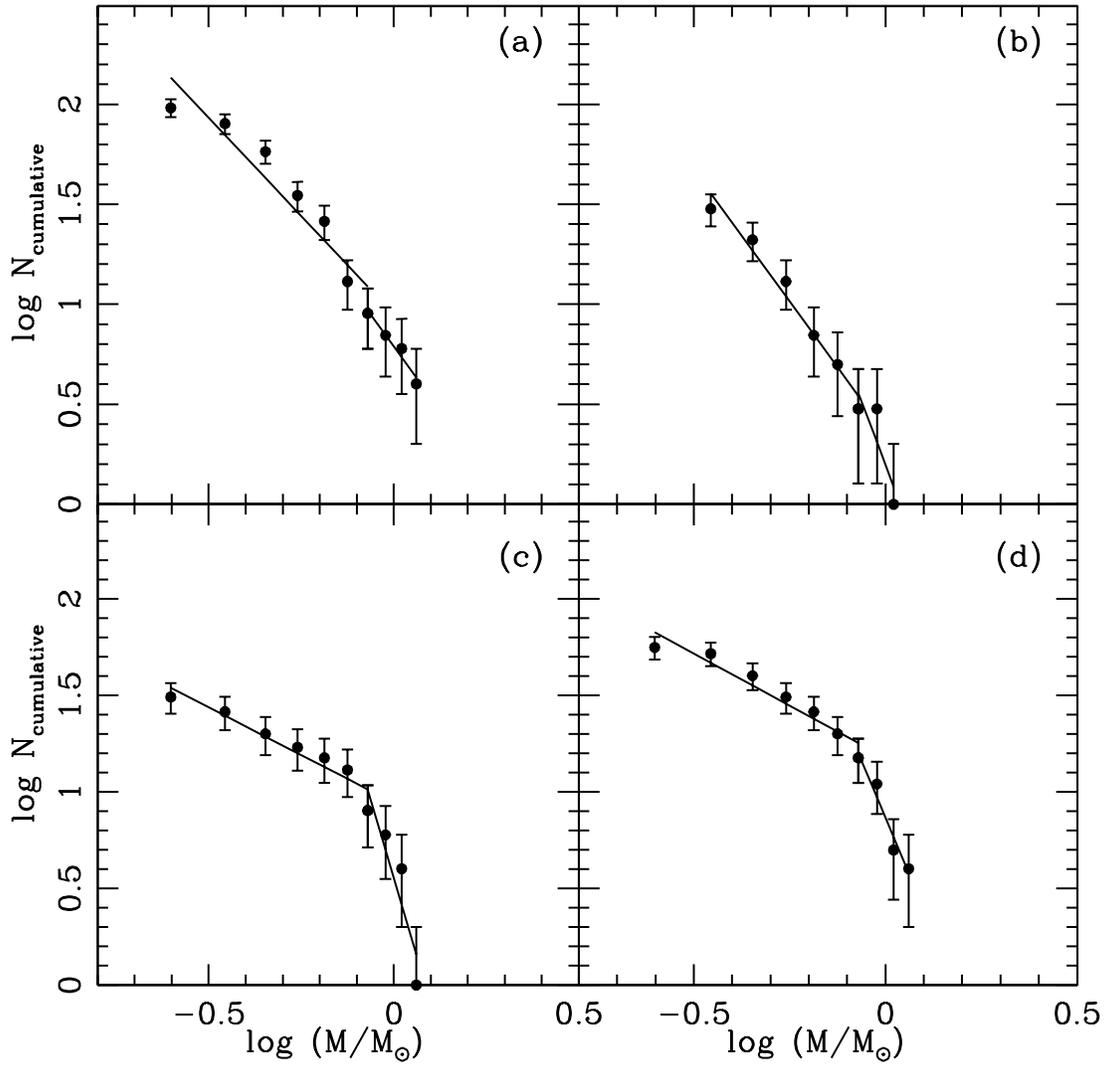}
\caption{CMFs of YSOs in the (a) cluster, (b) BRC NW, (c) BRC 13 and (d) BRC 14  regions. Error bars represent $\pm \sqrt {N}$ errors. }
\label{fig14}
\end{figure*}
\begin{figure*}
\centering
\includegraphics[scale = .5, trim = 5 5 5  5, clip]{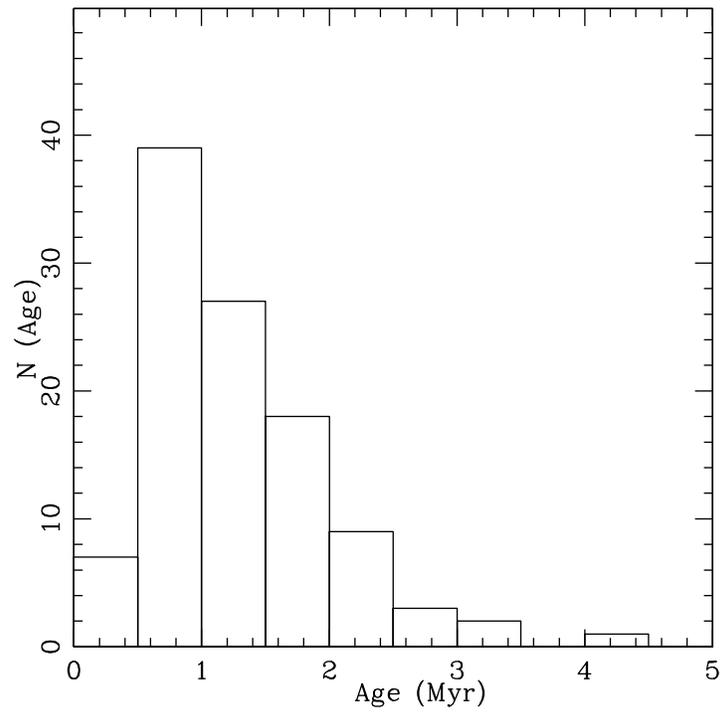}
\caption{Histogram of the age distribution for 
the YSOs in the cluster region.}
\label{fig15}
\end{figure*}
\begin{figure*}
\centering
\includegraphics[scale = .77, trim = 0 0 5  5, clip]{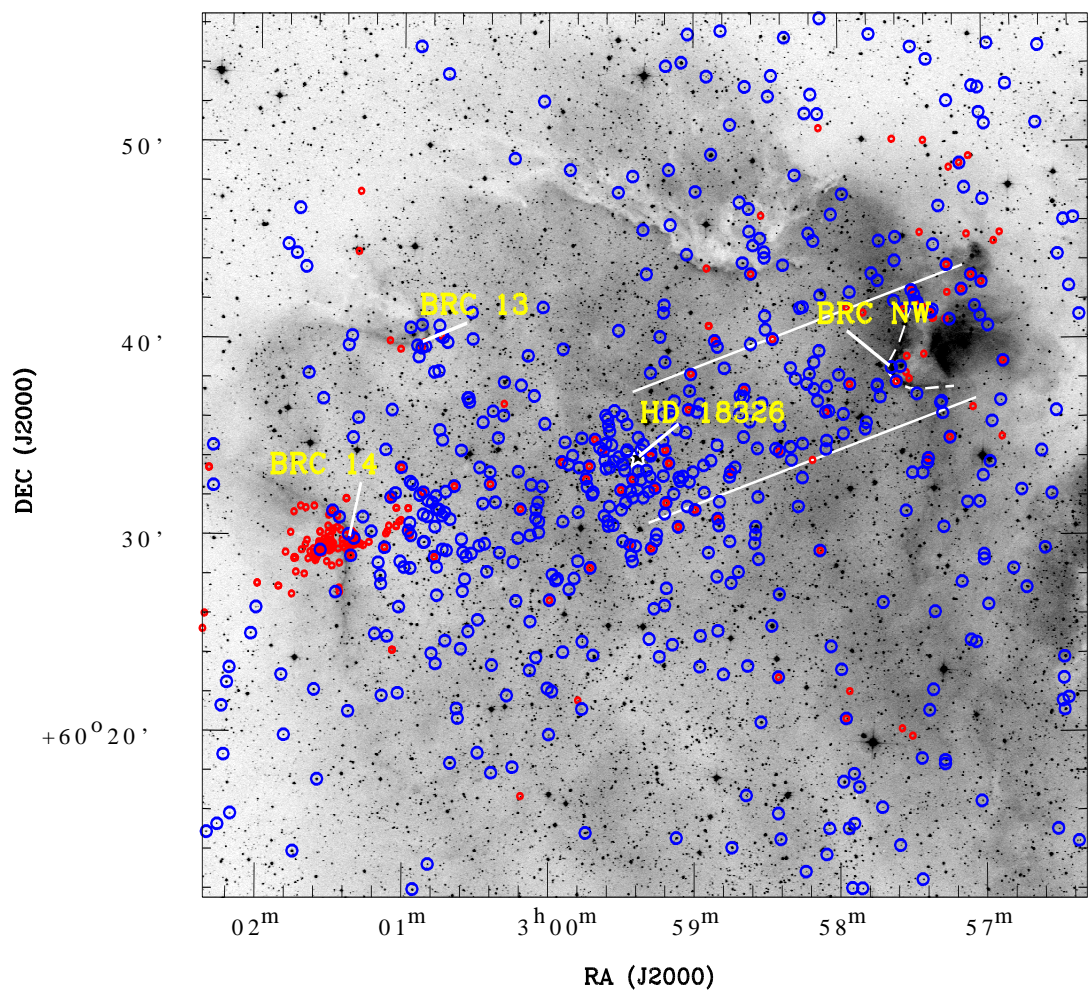}
\caption{ Global distribution of YSOs in the W5 E H{\sc ii} region. Small red circles show the 
location of NIR excess sources, whereas large blue circles (see the online version) 
show the YSOs identified using the $Spitzer$ observations (see text). }
\label{fig16}
\end{figure*} 
\begin{figure*}
\centering
\includegraphics[scale = .47, trim = 5 5 5  5, clip]{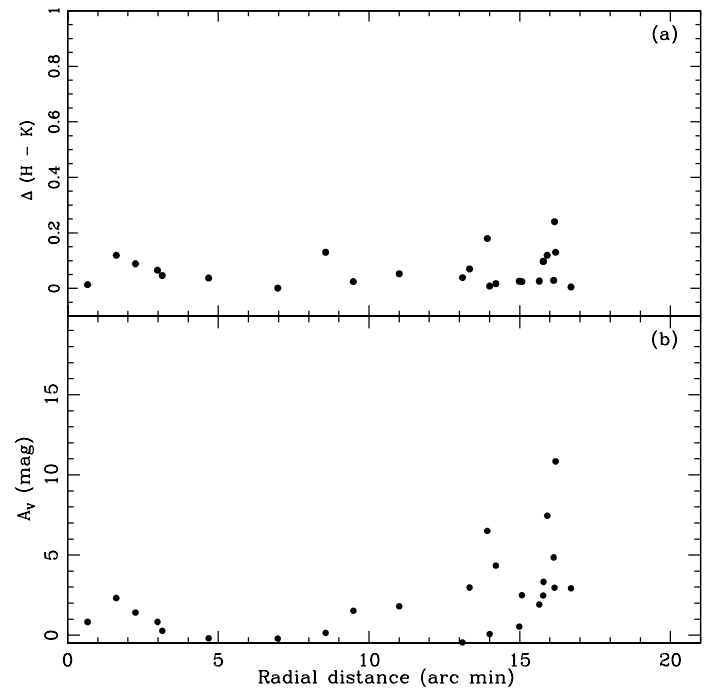}
\caption{ Variation of NIR excess $\Delta$({\it H-K}) (upper panel) and $A_V$ (lower panel) 
for the IR excess stars in the strip toward BRC NW as a function of distance
from the ionising source (HD 18326) of the W5 E H{\sc ii} region.}
\label{fig17}
\end{figure*}
\clearpage

\begin{figure*}
\centering
\includegraphics[scale = .45, trim = 5 5 5  5, clip]{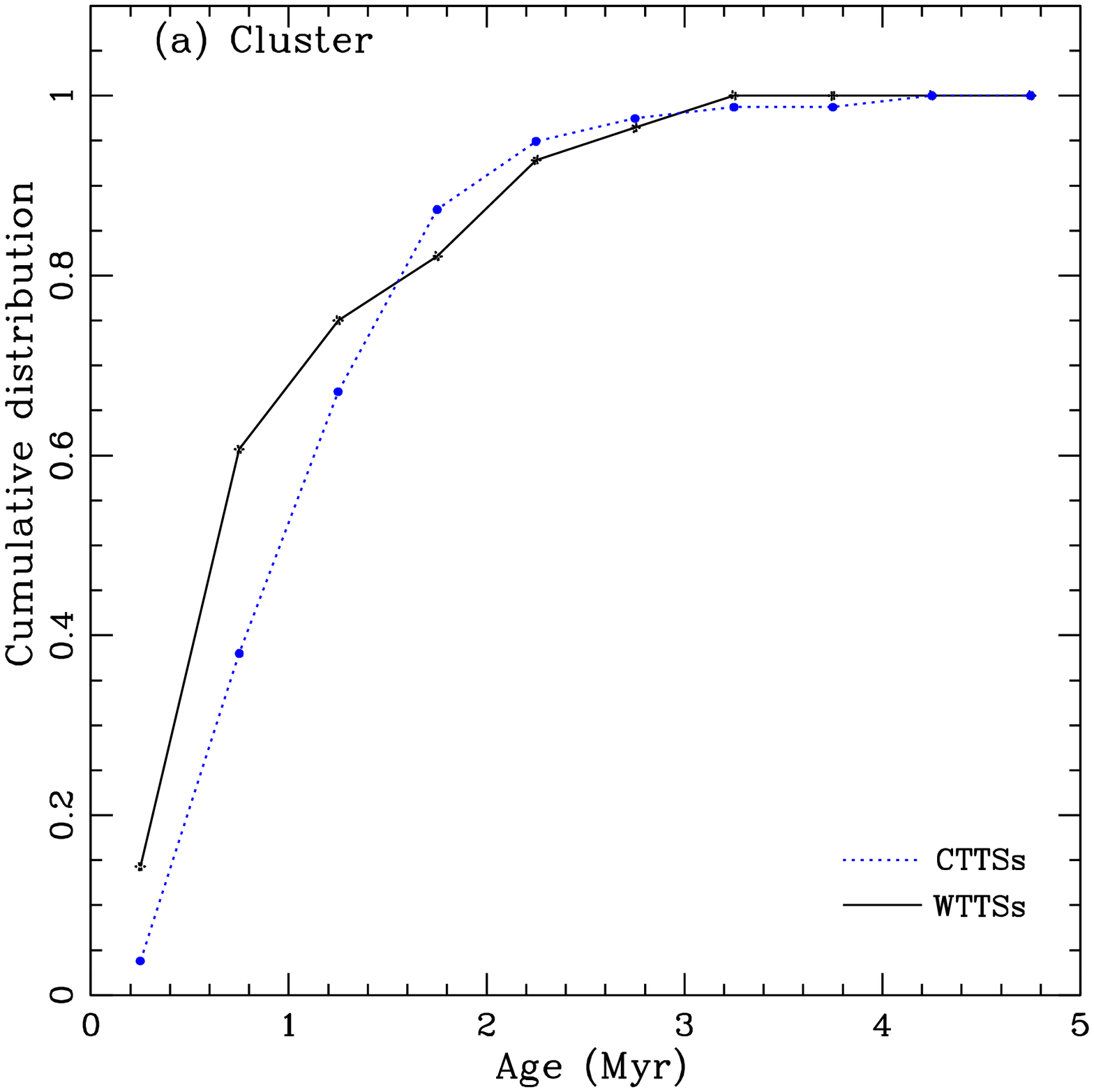}
\includegraphics[scale = .45, trim = 5 5 5  5, clip]{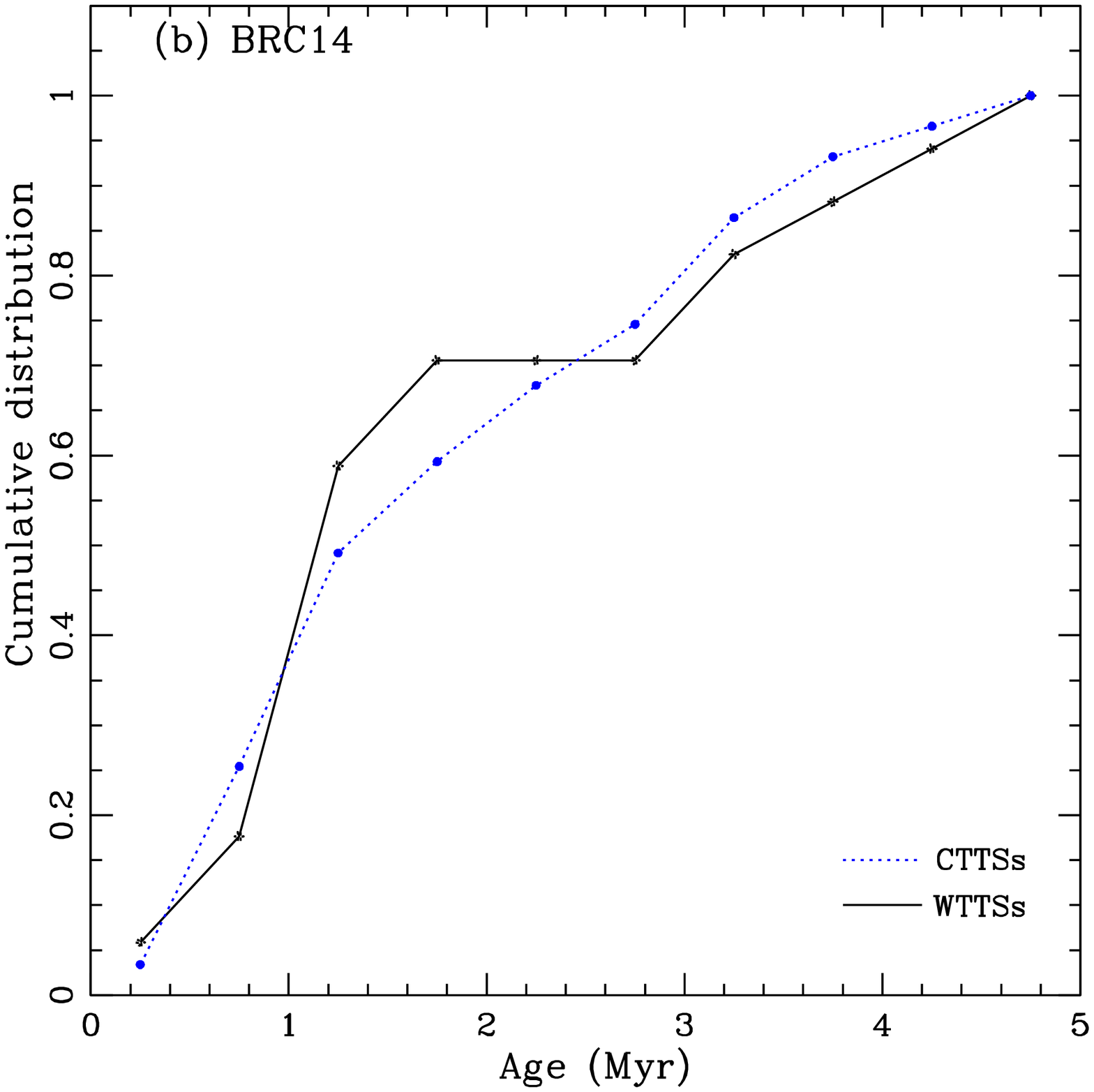}
\caption{Cumulative age distribution of Class II (CTTSs) and Class III (WTTSs) 
sources in the (a) cluster and (b) BRC 14 region.}
\label{fig19}
\end{figure*}
\bsp


\begin{thebibliography}{99}
\Large {\bibitem{b1}Armitage P.J., Clarke C.J., Palla F. 2003, MNRAS, 342, 1139
\bibitem{b2}Becker W., Fenkart R. 1971, A\&AS, 4, 241
\bibitem{b3}Bertoldi F. 1989, ApJ, 346, 735
\bibitem{b4}Bessell M.S., Brett J.M. 1988, PASP, 100, 1134 
\bibitem{b5}Bertout C., Siess L., Cabrit S. 2007, A\&A, 473, L21
\bibitem{b7}Chauhan Neelam, Pandey A.K., Ogura K., Ojha D.K., et al. 2009, MNRAS, 396, 694 (Paper II)
\bibitem{b8}Chini R., Wargau W. F., 1990, A\&A, 227, 5
\bibitem{b9}Chini R., Kr{\"u}egel E. 1983, A\&A, 117, 289
\bibitem{b10}Cohen J.G., Frogel J.A., Persson S.E., Ellias J.H. 1981, ApJ, 249, 481
\bibitem{b11}Conti P.S., Leep E. M. 1974, ApJ, 193, 113
\bibitem{b12}Cutri R.M., Skrutskie, M. F., van Dyk, S., et al., 2003, The IRSA 2MASS All Sky Point Source Catalog, NASA/IPAC Infrared Science Archive, http://irsa.ipac.caltech.edu/applications/Gator/
\bibitem{b15}Edwards S., Hartigan P., Ghandour L., \& Andrulis C. 1994, AJ, 108, 1056
\bibitem{b17}Georgelin Y.M., Georgelin Y.P. 1976, A\&A, 49, 57
\bibitem{b18}Girardi L., Bertelli G., Bressan A., Chiosi C., Groenewegen M.A.T., et al. 2002, 
A\&A, 391,195
\bibitem{b19}Gras-Vel{\'a}zquez A.,  Ray T. P. 2005, A\&A, 443, 541
\bibitem{b20}Hartmann L., Hewett R., Calvet N. 1994, ApJ, 426, 669
\bibitem{b21}Hartmann L. 2001, AJ, 121, 1030
\bibitem{b22}He Lida, Whittet D. C. B., Kilkenny D., Spencer Jones J. H., 1995, ApJS, 101, 335
\bibitem{b23}Hillenbrand L.A., Strom S.E., Vrba F.J., Keene J. 1992, ApJ, 397, 613
\bibitem{b23}Hillenbrand L.A. 2005, A Decade of Discovery: Planets Around Other Stars" STScI Symposium Series 19, ed. M. Livio, eprint arXiv:astro-ph/0511083
\bibitem{b23}Hillenbrand L.A., Bauermeister A. \& White R.J. 2008, in ASP Conf. Ser. 384, 14th Cambridge Workshop on Cool Stars, Stellar Systems, and the Sun, ed. G. van Belle (San Francisco, CA: ASP), 200
\bibitem{b24}Hillwig Todd C., Gies D.R., Bagnuolo W. G. Jr., Huang W., McSwain M. V., Wingert D.W. 2006, ApJ, 639, 1069
\bibitem{b25}Johnson H.L., Morgan W.W. 1953, ApJ, 117, 313
\bibitem{b26}Jose J., Pandey A.K., Ojha D.K., Ogura K., Chen W.P., Bhatt B.C., Ghosh S.K., 
Mito H., Maheswar G., Sharma S. 2008, MNRAS, 384, 1675
\bibitem{b27}Karr J.L., Martin P.G. 2003, ApJ 595, 900
\bibitem{b28} Kazarovets E.V., Durlevich O.V., Samus N.N. 1998, New Catalogue of 
Suspected Variable Stars, II/219, CDS, Strasbourg (NCSVS)
\bibitem{b29}Kenyon S., Hartmann L. 1995, ApJS, 101,117
\bibitem{b30}Kessel-Deynet O.,  Burkert A. 2003, MNRAS, 338, 545
\bibitem{b31}King I., 1962, AJ, 67, 471
\bibitem{b32}Koenig X.P., Allen L.E., Gutermuth R A., Hora J.L., Brunt C.M., Muzerolle J. 
2008, ApJ, 688, 1142 
\bibitem{b33}Kroupa P. 2001, MNRAS, 322, 231
\bibitem{b34}Kroupa P. 2002, Science, 295, 82
\bibitem{b37}Landolt A.U. 1992, AJ, 104, 340
\bibitem{b38} Lawson W.A., Fiegelson E.D., Huenemoerder D.P. 1996, MNRAS, 280, 1071
\bibitem{b39}Lefloch B., Lazareff B. 1995, A\&A, 301, 522
\bibitem{b42}Meyer M., Calvet N., Hillenbrand L.A. 1997, AJ, 114, 288
\bibitem{b43}Miao J., White G.J., Nelson R., Thompson M., Morgan L. 2006, MNRAS, 369, 143
\bibitem{b45}Mart{\'{\i}}n E.L. 1998, AJ, 115, 351
\bibitem{b46}Martins F., Plez B. 2006, A\&A, 457, 637
\bibitem{b47}Matsuyanagi I., Itoh Y., Sugitani K., Oasa Y., Mukai T., Tamura M. 2006, PASJ, 58, L29
\bibitem{b48}Morgan L.K., Thompson M.A., Urquhart J.S., White G.J. 2008, A\&A, 477, 557
\bibitem{b49}Muzerolle J., Calvet N., Hartmann L. 2001, ApJ, 550, 994
\bibitem{b50}Nakano M., Sugitani K., Niwa T., Itoh Y., Watanabe M. 2008, PASJ, 60, 739
\bibitem{b51}Neckel T., Chini R. 1981, A\&AS, 45, 451
\bibitem{b52}Niwa T., Tachihara K., Itoh Y., Oasa Y., et al. 2009, A\&A, 500, 1119 
\bibitem{b53}Ogura K., Sugitani K., Pickles A. 2002, AJ, 123, 2597
\bibitem{b54}Ogura K., Chauhan N., Pandey A.K., Bhatt B.C., Ojha D.K., Itoh Y. 2007, PASJ, 59, 199 (Paper I)
\bibitem{b55}Ojha D.K., Tamura M., Nakajima Y., et al. 2004a, ApJ, 608, 797
\bibitem{b56}Ojha D.K., Tamura M., Nakajima Y., et al. 2004b, ApJ, 616, 1042
\bibitem{b57}Palla F., Stahler S.W. 1999, ApJ, 525, 772
\bibitem{b58}Pandey A. K., Ogura K., Sekiguchi K., 2000, PASJ, 52, 847
\bibitem{b59}Pandey A.K., Nilakshi, Ogura K., Sagar R., Tarusawa K. 2001, A\&A, 374, 504
\bibitem{b60}Pandey A.K., Upadhyay K., Nakada Y., Ogura K., 2003, A\&A, 397, 191 
\bibitem{b61}Pandey A.K., Sharma S., Ogura K. 2006, MNRAS, 373, 255
\bibitem{b62}Pandey A.K., Sharma S., Ogura K., Ojha D.K., Chen W.P., Bhatt B.C., Ghosh S.K. 2008, MNRAS, 383, 1241
\bibitem{b63}Pandey et al. 2010, in preparation
\bibitem{b65}Robitaille T.P., Whitney B.A., Indebetouw R., Wood K., Denzmore P. 2006, ApJS, 167, 256
\bibitem{b66}Salpeter E.E. 1955, ApJ, 121, 161 
\bibitem{b67}Samal M.R., Pandey A.K., Ojha D.K., Ghosh S.K., Kulkarni V.K., Bhatt B.C. 2007, ApJ, 671, 555
Tamura M., Baba D., Sato S., Tsujimoto M. 2007, ApJ, 667, 963
\bibitem{b69}Schmidt-Kaler Th. 1982, Landolt-Bornstein, Vol. 2b, ed. K. 
Schaifers, H. H. Voigt, H. Landolt (Berlin: Springer), 19
\bibitem{b71}Siess L., Dufour E., Forestini M. 2000, A\&A, 358, 593
\bibitem{b72}Stetson P.B. 1987, PASP, 99, 191
\bibitem{b73}Stetson P.B. 1992, ASPC, 25, 297
\bibitem{b74}Sugitani K., Fukui Y., Ogura K. 1991, ApJS, 77, 59
\bibitem{b75}Sugitani K., Tamura M., Ogura K. 1995, ApJ, 455, L39
\bibitem{b76}Tapia M., Costero R., Echevarria J., Roth M., 1991, MNRAS, 253, 649
\bibitem{b77}Turner N.H., ten Brummelaar T.A., Roberts L.C., Mason, B.D., Hartkopf W.I., Gies D.R. 2008, AJ, 136, 554
\bibitem{b78}Walborn N.R. 1973, AJ, 78, 1067 
\bibitem{b79}Walter F.M., Brown A. Matthieu R.D., Meyer P.C., Vrba F.J. 1988, AJ, 96, 297
\bibitem{b80}Wegner W. 1993, AcA, 43, 209
\bibitem{b81}Winkler H. 1997, MNRAS, 287, 481}
\end{thebibliography}
\end{document}